\documentclass[ALICE,manyauthors]{cernphprep}
\usepackage[comma,square,numbers,sort&compress]{natbib}

\usepackage{lineno}
\usepackage{xspace}
\usepackage{hyperref}
\usepackage{upgreek}
\usepackage{mathrsfs}
\usepackage[T1]{fontenc}
\usepackage{orcidlink}

\newcommand{\krl}{\ensuremath{\kern-0.18em}}
\newcommand{\krr}{\ensuremath{\kern-0.09em}}
\newcommand{\tms}{\ensuremath{\kern-0.1em\times\kern-0.2em}}
\newcommand{\ptt}{\ensuremath{p_{\mathrm{T}}}\xspace}
\newcommand{\mT}{\ensuremath{m_{\mathrm{T}}}\xspace}

\newcommand{\pb}{Pb--Pb\xspace}
\newcommand{\ppb}{p--Pb\xspace}

\newcommand{\rs}[1][13~\TeV]{\ensuremath{\sqrt{s}=}~#1\xspace}

\newcommand{\rsnn}[1][5.02~\TeV]{\ensuremath{\sqrt{s_{\mathrm{NN}}}=}~#1\xspace}

\newcommand{\gvc}{\GeVc}
\newcommand{\gvcc}{\GeVmass}

\newcommand{\mvcc}{\MeVmass}

\newcommand{\pion}{\ensuremath{\pi}\xspace}
\newcommand{\pix}{\ensuremath{\pion^{\pm}}\xspace}

\newcommand{\km}{\ensuremath{\mathrm{K}^{-}}\xspace}

\newcommand{\kz}{\ensuremath{\mathrm{K}^{0}_{\mathrm{S}}}\xspace}

\newcommand{\la}{\ensuremath{\Lambda}\xspace}
\newcommand{\laa}{\ensuremath{\overline{\Lambda}}\xspace}
\newcommand{\xim}{\ensuremath{\Xi^{-}}\xspace}
\newcommand{\xip}{\ensuremath{\overline{\Xi}^{+}}\xspace}

\newcommand{\omx}{\ensuremath{\Omega}\xspace}

\newcommand{\oms}{\ensuremath{\Omega(2012)}\xspace}
\newcommand{\omsm}{\ensuremath{\Omega(2012)^{-}}\xspace}

\newcommand{\dd}{\ensuremath{\mathrm{d}}}
\newcommand{\mpt}{\ensuremath{\langle\ptt\rangle}\xspace}
\newcommand{\ddn}{\ensuremath{\dd^{2}N/(\dd\ptt\,\dd y)}\xspace}
\newcommand{\dnody}{\ensuremath{\dd N/\dd y}\xspace}
\newcommand{\dnodym}{\ensuremath{(\dd N/\dd y)_{\rm measured}}\xspace}

\newcommand{\dnc}{\avdndeta}

\newcommand{\dedx}{\ensuremath{\dd E\krl/\krr\dd x}\xspace}
\newcommand{\stpc}{\ensuremath{\sigma_{\mathrm{TPC}}}\xspace}

\newcommand{\mxk}{\ensuremath{M_{\Xi\mathrm{K}}}\xspace}
\newcommand{\vz}{\ensuremath{\mathrm{V}^{0}}\xspace}
\newcommand{\axe}{\ensuremath{A\!\times\!\varepsilon}\xspace}

\begin{document}
%

\newcommand{\pp}           {pp\xspace}
\newcommand{\ppbar}        {\mbox{$\mathrm {p\overline{p}}$}\xspace}
\newcommand{\XeXe}         {\mbox{Xe--Xe}\xspace}
\newcommand{\PbPb}         {\mbox{Pb--Pb}\xspace}
\newcommand{\pA}           {\mbox{pA}\xspace}
\newcommand{\pPb}          {\mbox{p--Pb}\xspace}
\newcommand{\AuAu}         {\mbox{Au--Au}\xspace}
\newcommand{\dAu}          {\mbox{d--Au}\xspace}

\newcommand{\s}            {\ensuremath{\sqrt{s}}\xspace}
\newcommand{\snn}          {\ensuremath{\sqrt{s_{\mathrm{NN}}}}\xspace}
\newcommand{\pt}           {\ensuremath{p_{\rm T}}\xspace}
\newcommand{\meanpt}       {$\langle p_{\mathrm{T}}\rangle$\xspace}
\newcommand{\ycms}         {\ensuremath{y_{\rm CMS}}\xspace}
\newcommand{\ylab}         {\ensuremath{y_{\rm lab}}\xspace}
\newcommand{\etarange}[1]  {\mbox{$\left | \eta \right |~<~#1$}}
\newcommand{\yrange}[1]    {\mbox{$\left | y \right |~<~#1$}}
\newcommand{\dndy}         {\ensuremath{\mathrm{d}N_\mathrm{ch}/\mathrm{d}y}\xspace}
\newcommand{\dndeta}       {\ensuremath{\mathrm{d}N_\mathrm{ch}/\mathrm{d}\eta}\xspace}
\newcommand{\avdndeta}     {\ensuremath{\langle\dndeta\rangle}\xspace}
\newcommand{\dNdy}         {\ensuremath{\mathrm{d}N_\mathrm{ch}/\mathrm{d}y}\xspace}
\newcommand{\Npart}        {\ensuremath{N_\mathrm{part}}\xspace}
\newcommand{\Ncoll}        {\ensuremath{N_\mathrm{coll}}\xspace}
\newcommand{\dEdx}         {\ensuremath{\textrm{d}E/\textrm{d}x}\xspace}
\newcommand{\RpPb}         {\ensuremath{R_{\rm pPb}}\xspace}

\newcommand{\nineH}        {$\sqrt{s}~=~0.9$~Te\kern-.1emV\xspace}
\newcommand{\seven}        {$\sqrt{s}~=~7$~Te\kern-.1emV\xspace}
\newcommand{\twoH}         {$\sqrt{s}~=~0.2$~Te\kern-.1emV\xspace}
\newcommand{\twosevensix}  {$\sqrt{s}~=~2.76$~Te\kern-.1emV\xspace}
\newcommand{\five}         {$\sqrt{s}~=~5.02$~Te\kern-.1emV\xspace}
\newcommand{\twosevensixnn}{$\sqrt{s_{\mathrm{NN}}}~=~2.76$~Te\kern-.1emV\xspace}
\newcommand{\fivenn}       {$\sqrt{s_{\mathrm{NN}}}~=~5.02$~Te\kern-.1emV\xspace}
\newcommand{\LT}           {L{\'e}vy-Tsallis\xspace}
\newcommand{\GeVc}         {Ge\kern-.1emV/$c$\xspace}
\newcommand{\MeVc}         {Me\kern-.1emV/$c$\xspace}
\newcommand{\TeV}          {Te\kern-.1emV\xspace}
\newcommand{\GeV}          {Ge\kern-.1emV\xspace}
\newcommand{\MeV}          {Me\kern-.1emV\xspace}
\newcommand{\GeVmass}      {Ge\kern-.2emV/$c^2$\xspace}
\newcommand{\MeVmass}      {Me\kern-.2emV/$c^2$\xspace}
\newcommand{\lumi}         {\ensuremath{\mathcal{L}}\xspace}

\newcommand{\ITS}          {\rm{ITS}\xspace}
\newcommand{\TOF}          {\rm{TOF}\xspace}
\newcommand{\ZDC}          {\rm{ZDC}\xspace}
\newcommand{\ZDCs}         {\rm{ZDCs}\xspace}
\newcommand{\ZNA}          {\rm{ZNA}\xspace}
\newcommand{\ZNC}          {\rm{ZNC}\xspace}
\newcommand{\SPD}          {\rm{SPD}\xspace}
\newcommand{\SDD}          {\rm{SDD}\xspace}
\newcommand{\SSD}          {\rm{SSD}\xspace}
\newcommand{\TPC}          {\rm{TPC}\xspace}
\newcommand{\TRD}          {\rm{TRD}\xspace}
\newcommand{\VZERO}        {\rm{V0}\xspace}
\newcommand{\VZEROA}       {\rm{V0A}\xspace}
\newcommand{\VZEROC}       {\rm{V0C}\xspace}
\newcommand{\Vdecay} 	   {\ensuremath{V^{0}}\xspace}

\newcommand{\ee}           {\ensuremath{e^{+}e^{-}}} 
\newcommand{\pip}          {\ensuremath{\pi^{+}}\xspace}
\newcommand{\pim}          {\ensuremath{\pi^{-}}\xspace}
\newcommand{\kap}          {\ensuremath{\rm{K}^{+}}\xspace}
\newcommand{\kam}          {\ensuremath{\rm{K}^{-}}\xspace}
\newcommand{\pbar}         {\ensuremath{\rm\overline{p}}\xspace}
\newcommand{\kzero}        {\ensuremath{{\rm K}^{0}_{\rm{S}}}\xspace}
\newcommand{\lmb}          {\ensuremath{\Lambda}\xspace}
\newcommand{\almb}         {\ensuremath{\overline{\Lambda}}\xspace}
\newcommand{\Om}           {\ensuremath{\Omega^-}\xspace}
\newcommand{\Mo}           {\ensuremath{\overline{\Omega}^+}\xspace}
\newcommand{\X}            {\ensuremath{\Xi^-}\xspace}
\newcommand{\Ix}           {\ensuremath{\overline{\Xi}^+}\xspace}
\newcommand{\Xis}          {\ensuremath{\Xi^{\pm}}\xspace}
\newcommand{\Oms}          {\ensuremath{\Omega^{\pm}}\xspace}
\newcommand{\degree}       {\ensuremath{^{\rm o}}\xspace}

\begin{titlepage}
\PHyear{2025}       
\PHnumber{027}      
\PHdate{18 February}  

\title{Observation of the $\Omega$(2012) baryon at the LHC}
\ShortTitle{Observation of $\Omega$(2012)}   

\Collaboration{ALICE Collaboration\thanks{See Appendix~\ref{app:collab} for the list of collaboration members}}
\ShortAuthor{ALICE Collaboration} 

\begin{abstract}
A signal consistent with the \oms baryon has been observed with a significance of $15\sigma$ in pp collisions at $\sqrt{s}=13$~\TeV at the LHC. In this paper, the analysis technique is described and measurements of the mass and width of the \oms are reported, along with the first measurement of its transverse-momentum spectrum and yield. This paper corroborates the observation by Belle of this excited \omx state and the observation that the \oms has a rather narrow width for a strongly decaying resonance. The yield measurement is combined with a statistical thermal model calculation of strange baryon yield ratios to obtain estimates of the $\omsm\rightarrow\Xi\overline{\rm K}$ branching ratios. These results will improve our understanding of the internal structure and mass spectrum of excited baryon states and serve as a baseline for searches regarding modifications of these properties in high-temperature media.

\end{abstract}
\end{titlepage}

\setcounter{page}{2} 


\section{Introduction}

In 2018, the Belle Collaboration reported the first observation of an excited $\Omega$ baryon state, the \oms, with a significance of $7.2\sigma$~\cite{Belle_Omega2012_2018}. Belle discovered the particle via the decay channels $\omsm\rightarrow\Xi^{0}\km$, $\omsm\rightarrow\Xi^{-}\kz$, and their charge conjugates. The particle was reported to have a mass of $[2012.4\pm0.7\;\mathrm{(stat.)\pm0.6\;(sys.)}]$~\mvcc and a width of $\left[6.4^{+2.5}_{-2.0}\;\mathrm{(stat.)\pm1.6\;(sys.)}\right]$~\mvcc. In a later study, Belle reported a mass of $[2012.5\pm0.7\;\mathrm{(stat.)\pm0.5\;(sys.)}]$~\mvcc~\cite{Belle_Omega2012_2025}. Belle concluded that the \omsm is most likely to have a spin-parity configuration of $\mathrm{J^{P}}=\frac{3}{2}^{-}$. This was partly based on theoretical studies of excited $\Omega$ baryon states~\cite{Engel_2013,CapstickIsgur_1986,Loering_2001,PervinRoberts_2008,FaustovGalkin_2015,Oh_2007}.
Also, the measured width of about 6~\MeV favors a $d$-wave decay of a $\mathrm{J}=\frac{3}{2}$ state, rather than a significantly broader $s$-wave decay of a $\mathrm{J}=\frac{1}{2}$ state. Subsequently, multiple theoretical studies have supported a $\mathrm{J^{P}}=\frac{3}{2}^{-}$ configuration~\cite{XiaoZhong_2018,Aliev_2018a,Aliev_2018b,Wang_2018,Polyakov_2019,Liu_2020,Arifi_2022,Wang_2023,LinZou_2018,Huang_2018,Lu_2020,Hudspith_2024}.

Measurements of such excited hadronic states provide a wealth of information relevant to several topics of interest in high-energy particle and nuclear physics. Studies of the basic properties of excited hadrons, including their masses, widths, quantum numbers, and decay branching ratios, provide important tests and inputs for theoretical models of hadron structure. Prior to the discovery of the \oms, predictions of excited $\Omega$ states were obtained using various theoretical approaches, including lattice gauge theory~\cite{Engel_2013,Liang_2016}, the quark model~\cite{Chao_1981,Kalman_1982,CapstickIsgur_1986,Loering_2001,Liu_2007,PervinRoberts_2008,An_2013,AnZou_2014,FaustovGalkin_2015}, and the Skyrme model~\cite{Oh_2007}. Following the discovery, multiple theoretical studies have been published that explore possible explanations of the structure of the \oms~\cite{XiaoZhong_2018,Aliev_2018a,Aliev_2018b,Wang_2018,Polyakov_2019,Liu_2020,Liu_2021,Arifi_2022,Zhong_2023,Wang_2023,Valderrama_2018,LinZou_2018,PavaoOset_2018,Huang_2018,GutscheLyubovitskij_2020,Lu_2020,Lin_2020,Ikeno_2020,Ikeno_2022,Lyu_2023}. Some studies support the conclusion that the \oms is a regular baryon~\cite{XiaoZhong_2018,Aliev_2018a,Aliev_2018b,Wang_2018,Polyakov_2019,Liu_2020,Liu_2021,Arifi_2022,Zhong_2023,Wang_2023}, possibly a $P$-wave excitation. Several works note that the mass of the \oms is just below the combined mass of the kaon and the $\Xi(1530)$ and explore the possibility that the \oms has a hadronic molecule component~\cite{Valderrama_2018,LinZou_2018,PavaoOset_2018,Huang_2018,GutscheLyubovitskij_2020,Lu_2020,Lin_2020,Ikeno_2020,Ikeno_2022,Lyu_2023}. Configurations explored in these studies include $\overline{\rm K}\Xi(1530)$ and couplings of that state to other states such as $\overline{\rm K}\Xi$, $\Omega\eta$, and the three-quark configuration. The molecular interpretation is consistent with the recent Belle measurement of the resonant three-body decay $\omsm\rightarrow\Xi(1530)\overline{\rm K}\rightarrow\Xi\pi\overline{\rm K}$~\cite{Belle_Omega2012_2025}.

A full understanding of the hadron chemistry of the matter produced in hadronic or nuclear collisions requires measurements of the abundances of common hadrons as well as rare, excited hadrons. To have an accurate description of hadron yields, it is necessary to understand the spectrum of excited hadronic states that decay (feed down) into lower-mass states. This includes measurements of the yields of the excited states as well as their decay modes and branching ratios into lower-mass hadrons. Complete and up-to-date lists of hadronic states are needed for implementations of the Hadron Resonance Gas model in the framework of statistical thermal models~\cite{Hagedorn,Dashen,Andronic,Flor}, as well as for hadronic transport models~\cite{UrQMD,Petersen_UrQMD_hydro,Werner_EPOS4} and hydrodynamical simulations~\cite{Schenke_hydro,Jeon_hydro}. Better agreement with experimental results and lattice QCD is achieved by calculations that use hadron lists with all states catalogued by the PDG, as opposed to lists that  are restricted to well-established hadrons~\cite{Alba,Alba_2020,Karthein}. On the other hand, simple quark models often overestimate the number of excited states~\cite{CapstickIsgur_1986,Ebert}. Thus, experimental confirmation of rare strange resonant states such as the \oms is crucial. Accurate measurements of the properties of this state will help to improve the understanding of the spectra of excited strange baryonic states that should be included in such calculations. Furthermore, measurements of the yields of excited $\Omega$ states in nucleus-nucleus, proton-nucleus, or high-multiplicity proton-proton collisions could provide another way to explore the details of strangeness enhancement, the increase in strangeness production with increasing system size or event activity~\cite{Rafelski_Mueller,Hamieh_strangeness_2000,WA97_strangeness_enhancement,STAR_multistrange_AuAu130,ALICE_multistrange_PbPb276,ALICE_strangeness_pp}. In addition, central heavy-ion collisions appear to produce a hadron gas phase with a lifetime of several fm/$c$~\cite{ALICE_Kstar_phi_PbPb5_2020,EPOS_resonances_PbPb}, during which re-scattering and regeneration processes can modify the yields of short-lived hadronic resonances~\cite{Bleicher_Stoecker,Bliecher_Aichelin,UrQMD_Bass,Torrieri_thermal}. Studies of how \oms yields evolve relative to the yields of the ground-state $\Omega$ could provide a new observable to improve our knowledge of the interactions in the hadronic phase and its lifetime. The number of \oms candidates found for this analysis does not allow further subdivision of the measured sample into multiplicity classes. The ongoing runs at the LHC promise to increase the available data sample by a factor of at least 50~\cite{ALICE_upgrade_2024}. In the future, this may enable a multiplicity-dependent study of \oms production in pp collisions, as well as measurements in \ppb and \pb collisions, allowing the \oms to be used in hadron chemistry studies.

Furthermore, recent lattice QCD calculations of baryonic resonance properties near the pseudo-critical temperature indicate that chiral symmetry restoration in the quark--gluon plasma might be verifiable through medium modification of the mass and width of parity partners~\cite{Aarts_2017,ALICE_review_2024}. In particular, it was shown in ref.~\cite{Aarts_2018_PRD} that although the mass of the positive-parity partner stays unchanged in the crossover region, the mass of the heavier negative-parity partner is modified towards the on-shell mass of the positive-parity partner. The negative-parity baryonic states can be difficult to measure, but if the \oms is indeed the $\mathrm{J^{P}}=\frac{3}{2}^{-}$ partner of the $\frac{3}{2}^{+}$ ground-state \omx, the postulated mass shift might be measurable for the \oms.
Since chiral symmetry restoration can manifest itself either in a mass shift or a width broadening~\cite{Sasaki_chiral_2018,Dominguez_chiral_2012}, the requirement in both cases is to measure mass and width across a wide range of system sizes, centralities, or final-state particle multiplicities. The first mass and width measurements of the \oms~\cite{Belle_Omega2012_2018} are for particles produced in $\mathrm{e^{-}e^{+}}\rightarrow\Upsilon$ annihilation events, a low-multiplicity environment. This paper presents measurements of the \oms mass and width in pp collisions with a high-multiplicity (HM) trigger. The mean charged-particle multiplicity at midrapidity for this data set is approximately 6 times larger than for inelastic pp collisions at the same energy~\cite{ALICE_Nch_pp13}. However, far higher multiplicity values (hundreds of times larger than in inelastic pp collisions) can be reached in nucleus-nucleus collisions~\cite{ALICE_Nch_PbPb5}, which produce larger and longer-lived volumes of matter in which the signatures of chiral symmetry restoration might be observed. The upgraded performance of the ALICE detector, mentioned above, may therefore also enable searches for mass and width modifications of the \oms in large collision systems.

A signal consistent with the \oms baryon has been observed by the ALICE Collaboration at the LHC. This paper describes the measurement of the $\omsm\rightarrow\xim\kz$ decay and its charge conjugate in pp collisions at \rs[13]~\TeV. Approximately 7200 candidates are observed. Results reported here include measurements of the mass and width of the \oms and the significance of the measured signal. The first measurement of the transverse-momentum (\ptt) spectrum and \ptt-integrated yield of the \oms are also reported. Finally, the measured yields are used along with a statistical model calculation to estimate the branching ratios of the $\omsm\rightarrow\Xi\overline{\rm K}$ decay channels. In the discussion that follows, the symbols \omx and \oms (i.e., without superscripts indicating charges or a bar to denote antibaryons) refer to the sums of particles and antiparticles. The symbol \omx without any mass indicates the weakly-decaying ground-state \omx baryon with a mass around 1672~\mvcc~\cite{PDG}.

\section{Analysis Procedure}

\subsection{Apparatus}

A full description of the ALICE detector is provided in ref.~\cite{ALICE_detector,ALICE_performance_2014}. The sub-detectors that are relevant to this study of the \oms are the V0 detectors~\cite{ALICE_VZERO}, the Inner Tracking System (ITS)~\cite{ALICE_ITS}, the Time Projection Chamber (TPC)~\cite{ALICE_TPC}, and the Time-of-Flight (TOF) detector~\cite{ALICE_TOF}. The V0 detectors, which provide triggering, are scintillator arrays that are located on either side of the center of the ALICE detector near the beamline; the V0A array covers the pseudorapidity range $2.8<\eta<5.1$, and the V0C array covers $-3.7<\eta<-1.7$. During the data-taking period for this study, the ITS consisted of six cylindrical layers of silicon detectors that were located between 3.9 and 43~cm from the beamline. The two innermost layers of the ITS are called the Silicon Pixel Detector (SPD). The TPC is a large-volume gas detector that encloses the region with a radius from about 85 to 250~cm from the beamline. Together, the TPC and ITS span the pseudorapidity range $|\eta|<0.9$. The ITS and TPC are used for the reconstruction of charged-particle trajectories and finding the primary collision vertex (PV). The SPD plays a role in rejecting pileup events. In addition, the TPC is used for particle identification (PID) via measurements of the specific energy loss \dedx in the TPC gas. The TOF is a cylindrical array of multigap resistive plate chambers located beyond the outer wall of the TPC. For this study, the TOF is used in the rejection of pileup, as discussed below.

\subsection{Data Set}

\begin{sloppypar}The \oms baryons are reconstructed in a set of HM-triggered pp collisions at \rs that were recorded over 2016--2018, during run 2 of the LHC. The trigger for inelastic pp collisions requires a coincidence of signals in the V0A and the V0C arrays. The HM trigger selects collisions which produced large signals in the V0 detectors, preferentially recording events with the highest charged-particle multiplicities, representing approximately 0--0.1\% of the visible cross section. The \oms signal is extracted from $1.04\!\times\!10^{9}$ HM-triggered events, representing an integrated luminosity of $36.6\pm2.0\;\mathrm{pb}^{-1}$~\cite{ALICE_luminosity_pp13,ALICE_ee_HF_pp13}. This data set has a mean charged-particle multiplicity density at midrapidity of $\dnc=31.53\pm0.28$~\cite{ALICE_Nch_2021}.\end{sloppypar}

Protons circulate around the LHC in groups called ``bunches.'' Pileup may arise both from multiple collisions in the same bunch crossing (``in-bunch pileup'') or from collisions in other bunch crossings that did not fire the trigger, but are still within the readout time interval for some components of the ALICE detector (``out-of-bunch pileup''). The procedure for removing background and pileup collisions is adopted from ref.~\cite{ALICE_strange_pp13}. Beam-induced background events are removed through use of timing information in the V0 detectors, and a selection on the correlation between clusters and tracklets in the SPD; see ref.~\cite{ALICE_performance_2014} for a detailed discussion. In-bunch pileup is removed by discarding events with multiple PVs reconstructed in the SPD. Out-of-bunch pileup events are removed through selection on multiplicity correlations in detectors that have different readout windows. Residual contributions to the \oms signal from out-of-bunch pileup are removed by requiring that at least one of the five decay product tracks of each \oms candidate is matched to a signal in the ITS or the TOF. This takes advantage of the short time resolutions for those detectors in comparison to the TPC, ensuring that the \oms candidates come from events that fired the trigger.

\subsection{Decay Reconstruction}

The \oms baryon is reconstructed via the $\omsm\rightarrow\xim\kz$ decay channel and its charge conjugate. The decay products are reconstructed via the decays $\xim\rightarrow\pim\Lambda\rightarrow\pim(\mathrm{p}\pim)$, $\xip\rightarrow\pip\overline{\Lambda}\rightarrow\pip(\overline{\mathrm{p}}\pip)$, and $\kz\rightarrow\pip\pim$ using the topological reconstruction techniques described in ref.~\cite{ALICE_strange_pp13}. The decay product tracks are required to pass the selection criteria in the first section of Tab.~\ref{table:sel} to ensure good track reconstruction quality and that tracks do not originate from the PV. These include selections on the track pseudorapidity, the number and fraction of readout pad rows in the ALICE TPC that were used to reconstruct the track, and the distance of closest approach (DCA) of the track to the PV. The final-state \pix and (anti)protons are identified via measurements in the TPC of their specific energy loss \dedx. It is required that \pix have a \dedx within 5\stpc of the expected value, while (anti)protons must have a \dedx within 3\stpc of the expected value, where \stpc is the TPC \dedx resolution. The expected \dedx values for each particle species are calculated from the Bethe-Bloch formula.

Using these \pix and (anti)proton tracks, displaced decay vertices are reconstructed and \kz and $\Xi$ candidates are identified by their decay topologies. \kz, \la, and \laa are reconstructed based on their ``\vz'' decay topology: two oppositely charged tracks originating from the same secondary vertex, which is displaced from the PV. \xim(\xip) are reconstructed based on their ``cascade'' decay topology: a \la(\laa) and a \pim(\pip) originating from a single decay vertex, displaced from the PV. Selections based on the decay topology are applied to improve the purity of the $\Xi$ and \kz candidates, and reduce contributions from random combinations of decay products (see Tab.~\ref{table:sel} and discussions in refs.~\cite{ALICE_strange_pp13,ALICE_strangeness_pp900}). These include selections on the DCA between the trajectories of various particles in the decay chain, the DCA between particle trajectories and the PV, the radial position of the displaced decay vertex, the particle lifetime (computed from the momentum, distance travelled, and mass hypothesis), and the difference between the measured invariant mass and the expected mass value from the PDG~\cite{PDG}. Selections are also applied on the cosine of the ``pointing angle,'' the angle between the momentum vector of the particle and its displacement vector (the vector connecting its production and decay points). Selection of \kz candidates also employs ``competing \vz rejection:'' the mass hypothesis for either \kz decay product is changed from pion to (anti)proton; if this results in a \vz mass within 4~\mvcc of the \la mass, the \vz is rejected as a \kz candidate. The aforementioned selection criteria were varied during the evaluation of the systematic uncertainties; their contribution to the systematic uncertainties is listed as ``decay product selection'' in Tab.~\ref{table:sys}. That table also lists the other systematic uncertainties which will be described in the following discussion. For each observable, the total systematic uncertainty is the sum in quadrature of the sources listed in Tab.~\ref{table:sys}.

\begin{table}
\caption{\label{table:sel} Criteria used to select $\vz$ and $\Xi$ candidates. ``DCA'' stands for ``distance of closest approach'' and ``PV'' denotes the primary collision vertex. The selection criteria for $\Lambda/\overline{\Lambda}$ are given in parentheses if different from the \kz selection criteria.}
\begin{center}
\begin{tabular}{ l l }
\hline
\textbf{Decay product track selection} & \textbf{Selection criterion}\\\hline
Pseudorapidity & $|\eta|<0.8$ \\
Number of crossed rows in TPC & $\geq70$ \\
$N_{\mathrm{crossed}}/N_{\mathrm{findable}}$ & $\geq0.8$ \\
DCA to PV & $>0.06$~cm \\
Pions: $(\dedx)_{\rm TPC}-(\dedx)_{\rm expected}$ & $<5\stpc$ \\
(Anti)protons: $(\dedx)_{\rm TPC}-(\dedx)_{\rm expected}$ & $<3\stpc$ \\\hline
\textbf{\vz selection} & \textbf{\kz $(\Lambda/\overline{\Lambda})$ selection criterion}\\\hline
Pseudorapidity & $|\eta|<0.8$ (no selection) \\
Transverse decay radius & $>0.9$~cm \\
Proper lifetime $(mcL/p)$ & $<20$~cm ($<40$~cm) \\
DCA between decay products & $<1$ $(<1.6)$~cm \\
DCA to PV & $<0.3$~cm ($>0.07$~cm) \\
Cosine of pointing angle & $>0.97$ \\
Mass tolerance & $<30$~\mvcc ($<6$~\mvcc) \\
Competing \vz rejection & $<4$~\mvcc (no selection) \\\hline
\textbf{$\Xi$ selection} & \textbf{Selection criterion}\\\hline
Pseudorapidity & $|\eta|<0.8$ \\
Transverse decay radius & $0.5<r<100$~cm \\
DCA between \pion and \la & $<1.4$~cm \\
Cosine of pointing angle & $>0.97$ \\
Mass tolerance & $<7$~\mvcc \\
\end{tabular}
\end{center}
\end{table}

\begin{table}
\caption{\label{table:sys} Contributions to the systematic uncertainties (in percent) of the results of this study. ``\dnodym'' refers to the \ptt-integrated \oms yield over the measured region $2.2<\ptt<10$~\gvc only, without \ptt extrapolation. The total systematic uncertainty is the sum in quadrature of the uncertainties from the separate sources. For \ddn, the percent uncertainties for the four measured \ptt intervals are averaged to obtain the quoted values for each source. ``NA'' indicates that an uncertainty does not apply to a given result. ``Negl.'' indicates that a the contribution is negligible.}
\begin{center}
\begin{tabular}{ c c c c c c c }
\hline
Source & Mass & Width & \ddn & \dnodym & \dnody & \mpt\\\hline
Decay product selection & 0.012 & 13.6 & 6.9 & 3.8 & 3.5 & 0.4 \\\hline
Combinatorial background & Negl. & Negl. & 3.4 & 1.5 & 0.7 & 0.5 \\\hline
Signal extraction & 0.003 & 30.0 & 9.3 & 5.3 & 3.0 & 0.2 \\\hline
Mass shift & 0.005 & NA & NA & NA & NA & NA \\\hline
Material budget & NA & NA & 2.1 & 3.0 & 2.9 & 0.1 \\\hline
Hadronic int. cross section & NA & NA & 0.1 & 0.2 & 0.1 & Negl. \\\hline
Acceptance $\times$ efficiency & NA & NA & 4.0 & 4.0 & 4.0 & Negl. \\\hline
Spectrum fit variations & NA & NA & 1.7 & 0.7 & $_{-11.5}^{+36.1}$ & $_{-12.0}^{+6.3}$ \\\hline
Correlated btwn. \ptt intervals & NA & NA & 5.1 & NA & NA & NA \\\hline
Total & 0.013 & 32.9 & 13.5 & 8.7 & $_{-13.4}^{+36.7}$ & $_{-12.0}^{+6.3}$ \\\hline
\end{tabular}
\end{center}
\end{table}

\begin{figure*}
\begin{centering}
\includegraphics[width=19pc]{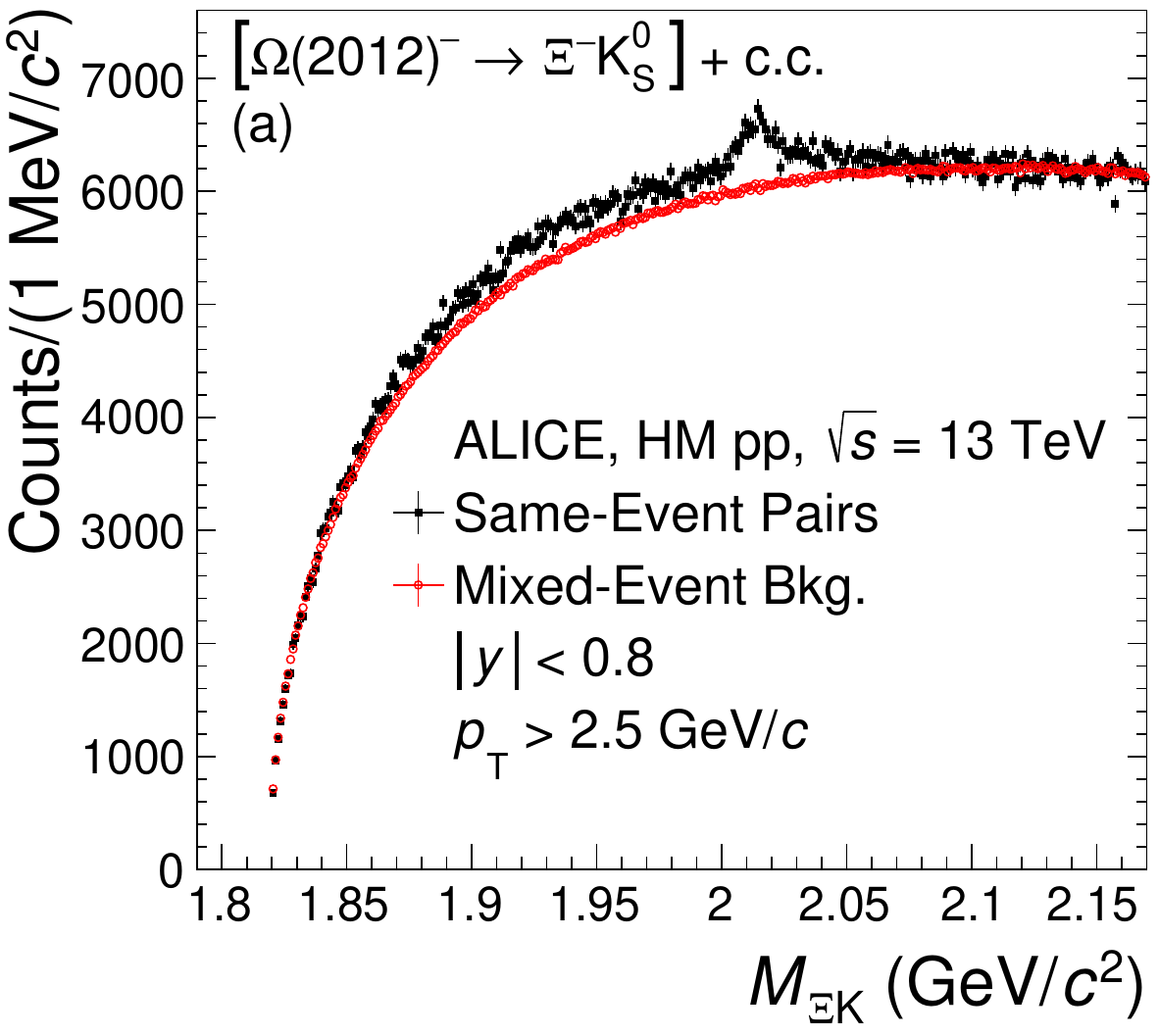}\\
\end{centering}
\includegraphics[width=19pc]{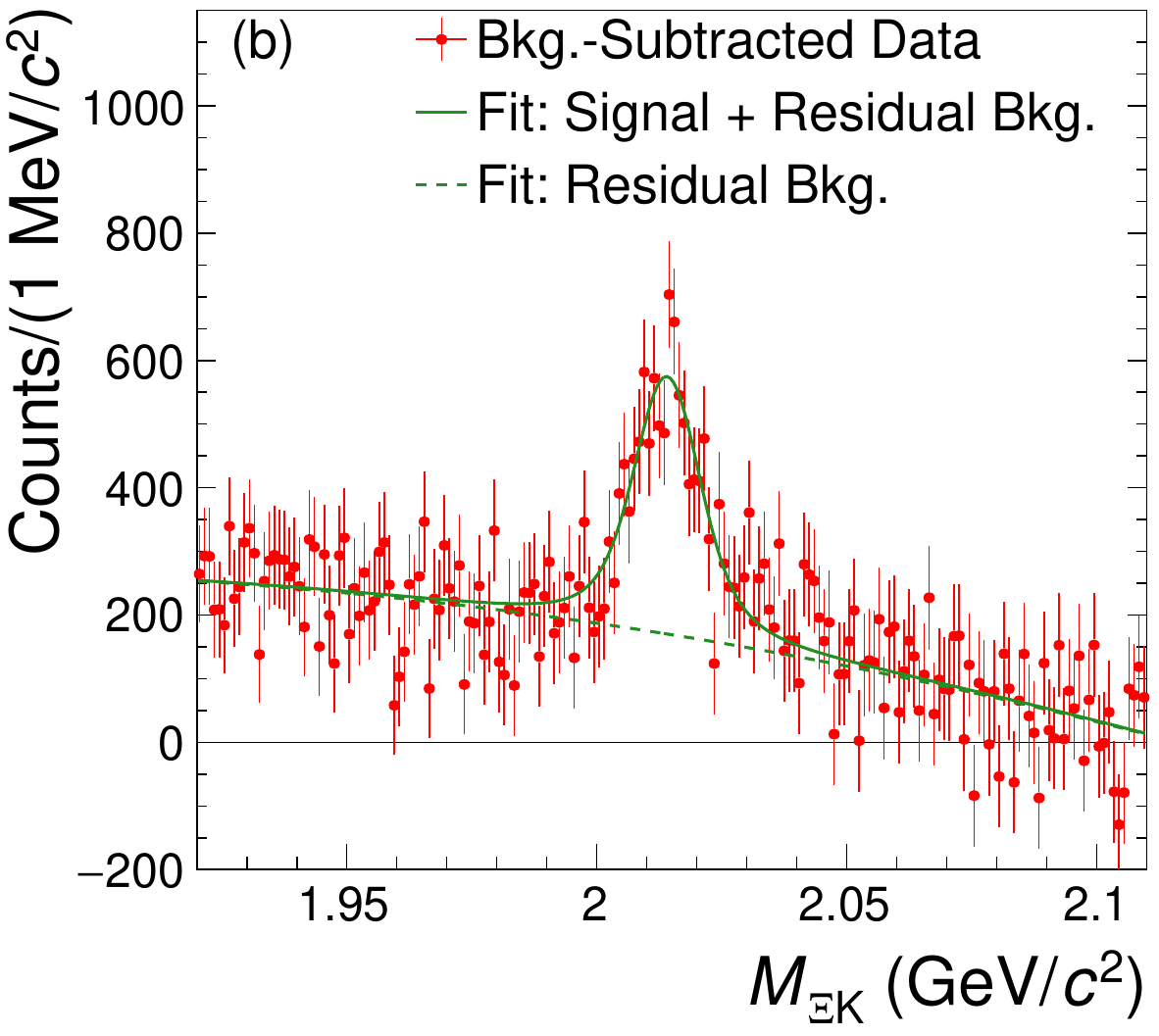}
\includegraphics[width=19pc]{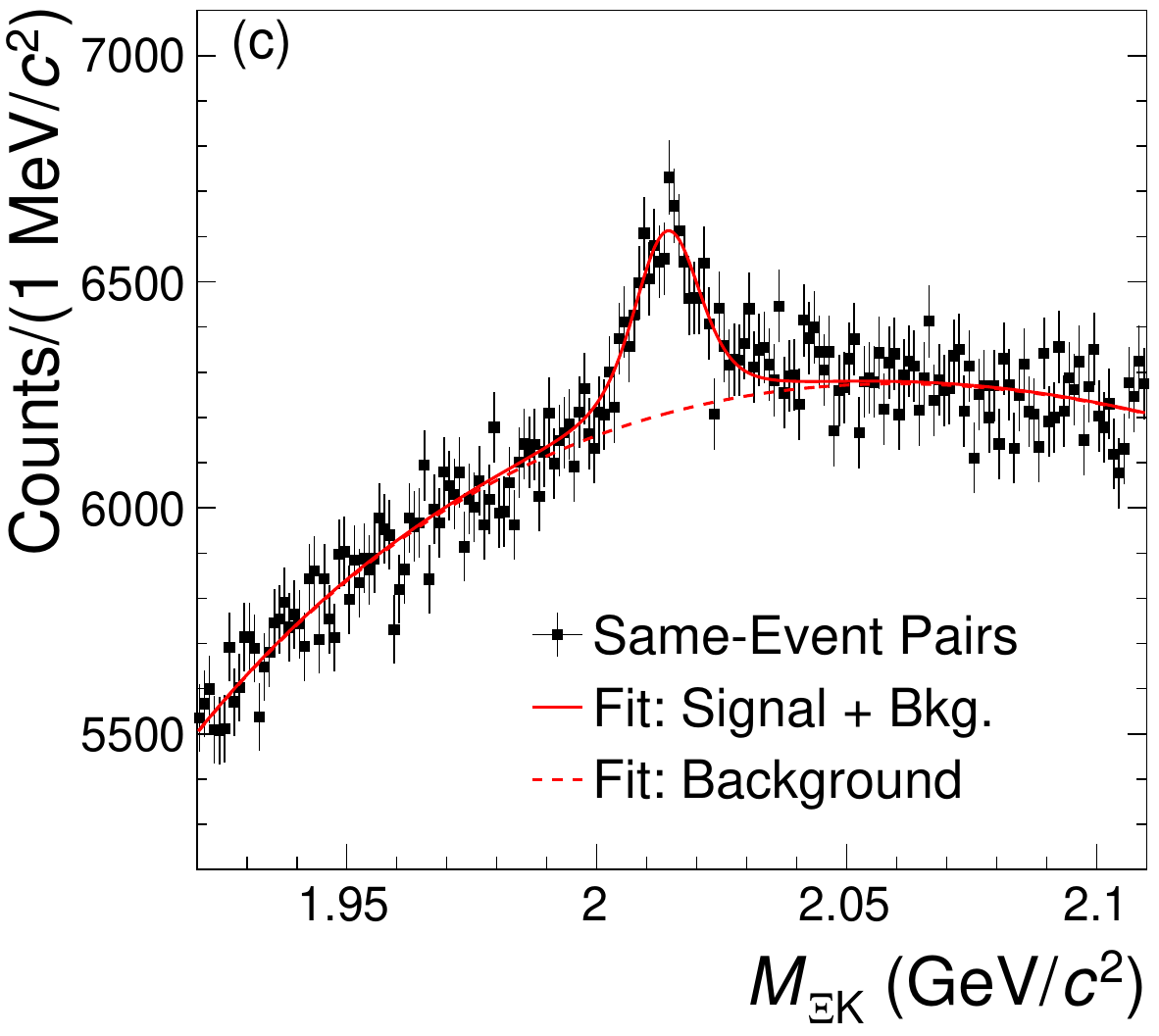}
\caption{\label{fig:mass}(Color online) Invariant-mass distributions of $\Xi\kz$ pairs with $\ptt>2.5$~\gvc and $|y|<0.8$ in HM-triggered pp collisions at \rs[13]~\TeV. (a) Same-event distribution plotted with normalized mixed-event combinatorial background. (b) Background-subtracted distribution fitted with residual background polynomial and Voigt peak. (c) Same-event distribution fitted with summed background function and Voigt peak.}
\end{figure*}

Each $\Xi$ candidate is then paired with each \kz candidate to obtain an invariant-mass distribution for $\Xi$\kz pairs. The $\Xi$ and \kz are assigned their respective average mass values from the PDG~\cite{PDG}. Each pair is required to have a rapidity within the range $|y|<0.8$. Figure~\ref{fig:mass}a shows the $\Xi$\kz invariant-mass distributions used for the mass and width measurements; the signal-to-background ratio is 0.033. Monte Carlo simulations indicate that the product of the acceptance and efficiency for the \oms decreases with decreasing \ptt and falls below 1\% for $\ptt<2$~\gvc (see Fig.~\ref{fig:eff} and the discussion below). The \oms yield is measured in four \ptt intervals covering the range $2.2<\ptt<10$~\gvc, with the low-\ptt region excluded due to the lack of a significant signal. For the measurements of the mass and width of the \oms, the transverse momentum is taken to be $2.5<\ptt<10$~\gvc; the larger minimum \ptt value for the mass/width study is chosen to obtain a larger signal-to-background ratio.
The number of $\Xi\kz$ pairs that satisfy all selection criteria and have an invariant mass \mxk in the range $2.00<\mxk<2.03$~\gvcc is small compared to the number of collisions: on average, $\sim2\times10^{-4}$ such \oms candidates are reconstructed per event. It is unlikely that a single event will have more than one reconstructed \oms candidate. However, if multiple candidates are identified in an event, all are kept and contribute to the invariant-mass distribution.

Two complementary methods are used to describe the background distribution. First, the event-mixing technique is employed. Each \kz candidate is paired with $\Xi$ candidates from 10 other events from the same data sample. For each pair of events that are mixed, the positions of the PVs along the beam axis are required to differ by less than 1~cm. The mixed-event background is scaled so that its integral is equal to the integral of the same-event distribution in the invariant-mass range $2.10<\mxk<2.13$~\gvcc. The normalization region is varied during the evaluation of systematic uncertainties. As can be seen in Fig.~\ref{fig:mass}a, this mixed-event background can roughly describe the shape of the combinatorial background, but a residual background remains. This background includes contributions from jets, decays of other particles in which one or more products are misidentified or missed, and uncorrelated combinations of particles that mimic the \vz or cascade topology. The mixed-event background distribution is subtracted from the same-event distribution and the result is fitted. The fit function consists of a function that describes the peak (described below), added to a second-order polynomial, which parameterizes the residual background.

In the second method used to describe the combinatorial background, the same-event invariant-mass distribution is fitted directly (without subtracting the mixed-event background). The fit function consists of a function that describes the peak (described below), added to a background function of the form

\begin{equation}
B(\mxk)=b_{0}+b_{1}\mxk+b_{2}\mxk^{2}+b_{3}\sqrt{\mxk-M_{\Xi}-M_{\rm K}},
\end{equation}

where $b_{0}$, $b_{1}$, $b_{2}$, and $b_{3}$ are free parameters. $M_{\Xi}$ and $M_{\rm K}$ are the average mass values of the \xim and $\mathrm{K}^{0}$ from the PDG~\cite{PDG}. The addition of the square-root term assists in describing the greater curvature on the low-mass side of the \oms peak; a similar term was used in the $\upphi$-meson studies described in refs.~\cite{ALICE_LF_pp13,ALICE_Kstar_phi_pp13} for the same reason. The fitting procedure for these two methods is illustrated in Fig.~\ref{fig:mass}b and c. The results obtained using the two methods are compatible within several percent. They are compared during the calculation of the systematic uncertainties (listed as ``combinatorial background'' in Tab.~\ref{table:sys}), and their arithmetic mean is used to calculate the mass, width, and yield values.

In all cases, a Voigt function is used to parameterize the signal component. This function has the form given in Eq.~\ref{eq:voigt} below. It is the convolution of a Breit-Wigner distribution and a Gaussian (which describes the mass resolution).

\begin{equation}
\label{eq:voigt}
P(\mxk)=\frac{A\Gamma}{(2\pi)^{3/2}\sigma}\!\int\limits_{-\infty}\limits^{\infty}\exp\left[-\frac{\left(\mxk-m^{\prime}\right)^{2}}{2\sigma^{2}}\right]\times\frac{1}{\left(m^{\prime}-\mu\right)^{2}+\Gamma^{2}/4}\dd m^{\prime}.
\end{equation}

Here, $A$ is a scale factor, $\Gamma$ is the Breit-Wigner width parameter, $\sigma$ is the Gaussian resolution parameter, $\mu$ is the most probable value of the distribution, and $m^{\prime}$ is the variable of integration. Monte Carlo simulations (described below) indicate that the $\sigma$ parameter of the Voigt peak is approximately 6~\mvcc, increasing with the transverse momentum of the decaying \oms. This is of the same order of magnitude as the width $\Gamma$ of the resonance, which motivates the choice of the Voigt function to describe the peak. The resolution parameter $\sigma$ is fixed to the value extracted from the aforementioned simulations for each \ptt bin. In the evaluation of the systematic uncertainties, $\sigma$ is fixed to the lowest and highest simulation values (described below), and alternatively, the Breit-Wigner width parameter $\Gamma$ of the Voigt peak is fixed to the Belle value~\cite{Belle_Omega2012_2018}. The invariant-mass distribution is fitted over the range $1.92<\mxk<2.11$~\gvcc, with several alternate intervals used during the evaluation of the systematic uncertainties. The mass and width are extracted from these fits. The simulations indicate that, due to detector effects in the \oms reconstruction procedure, the reconstructed \oms mass is $0.48\pm0.11$~\mvcc larger than the generated (initial) mass. The measured \oms mass is corrected to account for this, and the associated uncertainty is listed as ``mass shift'' in Tab.~\ref{table:sys}.

The yield is calculated by subtracting the integral of the background function from integral of the same-event invariant-mass distribution over the range $2.00<\mxk<2.03$~\gvcc. The yield outside this integration range is computed by integrating the tails of the peak fit function. This accounts for about 15\% of the peak area and is added to the yield as a correction. Alternatively, the yield can be extracted directly from the fit function, but the resulting variations in the yields are well within the systematic uncertainties. The systematic uncertainty due to variations in how the Voigt function $\sigma$ and $\Gamma$ parameters are fixed, fitting range, and invariant-mass bin width is listed as ``signal extraction'' in Tab.~\ref{table:sys}.

\subsection{Parameterized Spectrum}
\label{sec:parameterized_spectrum}

Several components of this analysis (discussed in subsequent sections) require a realistic estimate of the shape of the \oms \ptt spectrum for HM pp collisions at \rs. In particular, the spectral shape is needed to properly weight the calculations of the acceptance and efficiency, mass resolution, and mean mass shift. A realistic \ptt spectral shape is also needed to extrapolate the \oms yield to the unmeasured \ptt region. However, since this paper reports the first measurement of the \oms \ptt spectrum, an independent estimate of the spectral shape is needed.

To derive such a spectrum, the following procedure is used. First, the measured $\Omega$ \ptt spectra in pp collisions at \rs~\cite{ALICE_strange_pp13} in different charged-particle multiplicity classes are fitted with a L\'{e}vy-Tsallis function~\cite{Tsallis,Wilk_Tsallis_2000,STAR_strange_pp_2007}:

\begin{equation}
\label{eq:levy}
\frac{\dd^{2}N}{\dd\ptt\;\dd y}=\ptt\times\frac{A(n-1)(n-2)}{nC\left[nC+M(n-2)\right]}\left[1+\frac{\sqrt{\ptt^{2}+M^{2}}-M}{nC}\right]^{-n}.
\end{equation}

Here, $A$, $C$, and $n$ are free parameters, while $M$ is fixed to the $\Omega$ mass. The  parameters $C$ and $n$ are observed to increase approximately linearly with increasing \dnc, the mean charged-particle multiplicity density at midrapidity. Next, linear functions are used to extrapolate the parameters $C$ and $n$ to $\dnc=31.53$, the mean multiplicity for the data sample in which this \oms study is performed~\cite{ALICE_Nch_2021}. Then an \mT-scaling procedure is used to convert the ground-state $\Omega$ \ptt spectrum to an estimated \ptt spectrum for the \oms. The \mT-scaling hypothesis assumes that the invariant cross section

\begin{equation*}
\label{eq:xsection}
E\frac{\dd^{3}\sigma}{\dd^{3}p}=\frac{\dd^{3}\sigma}{\dd\varphi\;\dd y\;\ptt\;\dd\ptt}=\frac{\dd^{3}\sigma}{\dd\varphi\;\dd y\;\mT\;\dd\mT}
\end{equation*}

has the same shape as a function of $\mT\equiv\sqrt{\ptt^{2}+M^{2}}$ for all baryons. In refs.~\cite{STAR_strange_pp_2007,ALICE_LF_pp13}, baryon spectra measured at STAR and ALICE in inelastic pp collisions appear to follow common trends as functions of \mT, consistent with this hypothesis. The \oms \ptt spectrum is taken to have the same form as Eq.~\ref{eq:levy} with the substitution $\sqrt{\ptt^{2}+M^{2}}\rightarrow\sqrt{\ptt^{2}+M_{\oms}^{2}}$. The remaining instances of the mass parameter $M$ are fixed to the ground-state $\Omega$ mass, and the parameters $C$ and $n$ are fixed to the values derived from the ground-state $\Omega$ \ptt spectra as discussed above. It has been verified that the ground-state $\Omega$ \ptt spectrum in the highest measured multiplicity class in pp collisions at \rs can be reproduced within its uncertainties through a similar \mT-scaling procedure, starting from the $\Xi^{-}$ \ptt spectrum.
The parameterized \oms spectrum is shown in Fig.~\ref{fig:spectrum}, along with the measured \ptt spectrum (discussed below).
Alternate parameterized \ptt spectra are obtained through \mT scaling of other baryon \ptt spectra, using an \mT-scaled \omx spectrum without multiplicity extrapolation, and extrapolating the L\'{e}vy-Tsallis parameters to a larger multiplicity value, $\dnc=35.82$ (corresponding approximately to the 0-0.01\% of the visible cross section with the highest multiplicities)~\cite{ALICE_Nch_2021}. The results of this study are found to be largely insensitive to the choice of the \ptt spectral shape. As discussed below, the alternate \ptt spectra are used to calculate systematic uncertainties for the affected observables.

\subsection{Quantities Derived from Simulations}

To evaluate the resolution parameter of the Voigt function, the mass shift, and the product of the acceptance and reconstruction efficiency for the \oms, simulations of \oms baryons embedded into PYTHIA~8~\cite{PYTHIA8,PYTHIA8_Monash} events were produced. The masses of the embedded \oms baryons followed a Breit-Wigner distribution with the width and most probable value set to the values measured by Belle~\cite{Belle_Omega2012_2018}. The simulated final-state particles were propagated through a GEANT~4~\cite{GEANT4} simulation of the ALICE detector and the same reconstruction procedures that were used for the real data.

\begin{figure*}
\centering
\includegraphics[width=20pc]{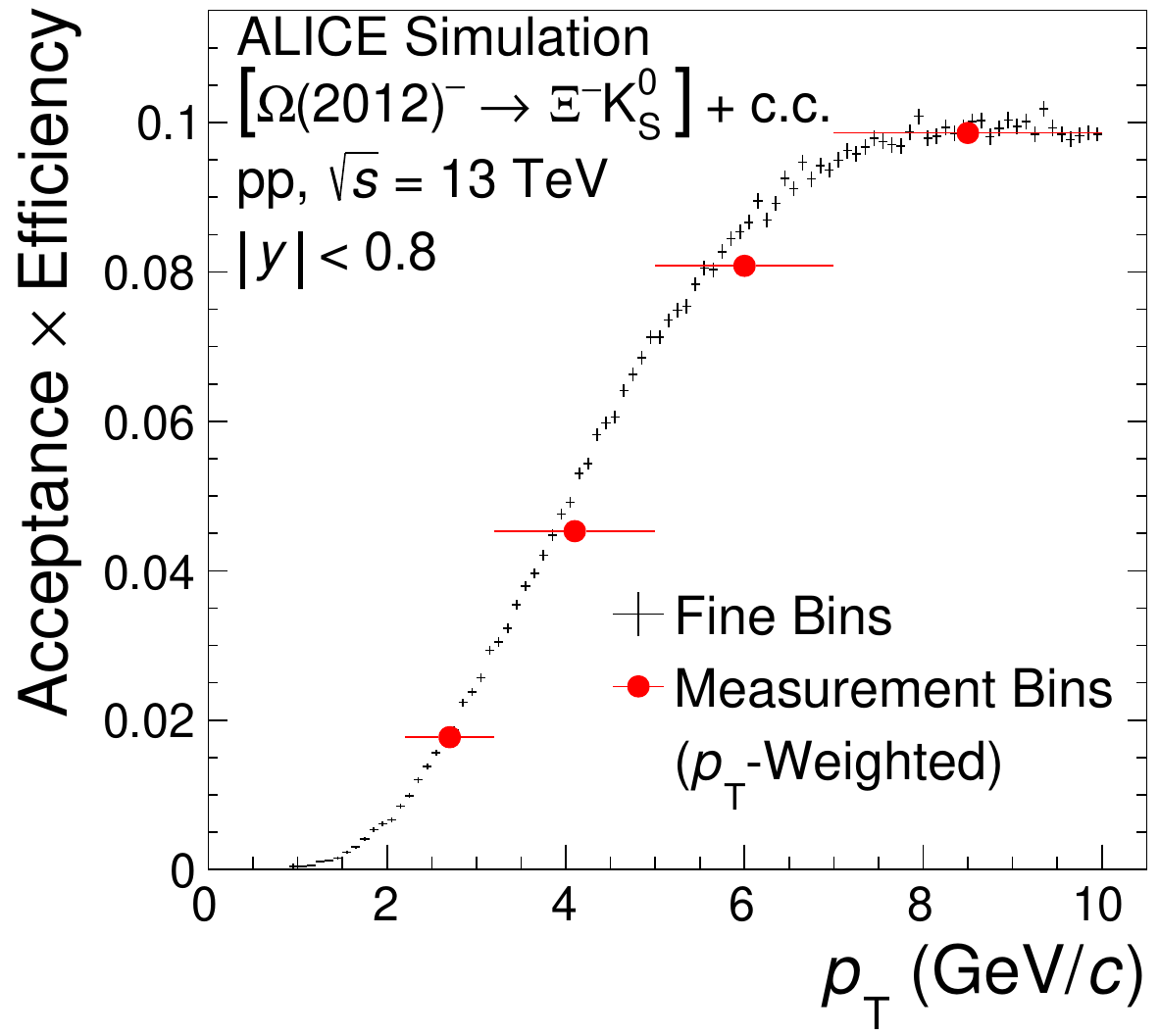}
\caption{\label{fig:eff}(Color online) Product of the acceptance and efficiency (\axe) for \oms baryons within the rapidity range $|y|<0.8$ in pp collisions at \rs[13]~\TeV. The value of \axe is shown in the same wide \ptt bins as used in the measurement of the \ptt spectrum in the real data (after the \ptt-weighting procedure) and also in fine \ptt bins. Only statistical uncertainties are shown.}
\end{figure*}

The product of the acceptance and the efficiency, denoted \axe, is the fraction of the \oms originally generated in the simulation that go on to be successfully reconstructed, with all decay products satisfying the selection criteria. This quantity is shown in Fig.~\ref{fig:eff} as a function of \ptt for the rapidity range $|y|<0.8$. It is observed that \axe for the \oms is independent of the particle's mass, allowing for the use of an unmodified Voigt function to describe the peak. The \oms yield in each \ptt interval is corrected by \axe to obtain the \ptt spectrum shown in Fig.~\ref{fig:spectrum}. In order to account for possible multiplicity mismatches between the real data and the simulation, a constant 4\% systematic uncertainty is assumed to be associated with \axe, contributing to the systematic uncertainties for the \ptt-differential and \ptt-integrated \oms yields. This value is derived from~\cite{ALICE_strange_pp13,ALICE_Xi_Omega_pPb5,ALICE_piKpLambda_pPb5}, in which a constant 2\% systematic uncertainty is applied to \kz, $\Lambda$, $\Xi$, and $\Omega$ yield measurements in pp and \ppb collisions to account for possible multiplicity dependence of the efficiencies. The \oms decays to a \kz and a $\Xi$, so the linear sum of the two \axe uncertainties for those decay products is assumed for this study. This uncertainty is listed as ``acceptance$\times$efficiency'' in Tab.~\ref{table:sys}. The statistical uncertainty of \axe is negligible for the four wide \ptt bins in which the yield is measured. The simulated \oms were uniformly distributed in transverse momentum over the range \mbox{$1<\ptt<10$~\gvc}. Therefore, during the \axe calculation, the simulated \oms are weighted based on their \ptt, using the parameterized \ptt spectrum described in Sec.~\ref{sec:parameterized_spectrum}. The \axe, and therefore the \oms yield, has only a small dependence on the parameterized \ptt spectrum. The ``spectrum fit variations'' systematic uncertainty listed in Tab.~\ref{table:sys} is estimated from the variations in the \ptt-differential \oms yield when alternate parameterized \ptt spectra are used to calculate \axe.

The mass resolution and mean mass shift (average difference between the generated and reconstructed masses) are also extracted from these simulations. The resolution is used in the Voigt fits described above, while the mass shift is used to correct the measured mass. Multiple techniques are used to extract these quantities, which are used to obtain their systematic uncertainties. First, the reconstructed mass distribution is fitted with a Voigt function. Second, the difference $\Delta M\equiv M_{\rm reconstructed}-M_{\rm generated}$ is calculated for each simulated \oms, and the resolution and mean mass shift are extracted from the $\Delta M$ distribution. The extracted parameters are weighted to account for the expected \ptt distribution for reconstructed \oms baryons, which is the product of the parameterized \ptt spectrum, the acceptance, and the efficiency (more details on these factors are presented below). The systematic uncertainties of the resolution and mass shift also account for the use of alternate parameterized \ptt spectra in this weighting procedure.

\section{Results}

\subsection{Mass and Width}

The mass and width of the \oms are observed to be

\begin{equation}
\label{eq:mass}
M_{\oms}=\left[2013.35\pm0.57\;(\mathrm{stat.})\pm0.27\;(\mathrm{sys.})\right]\;\mathrm{Me}\kern-.2em\mathrm{V}\!/\!c^{2},
\end{equation}

\begin{equation}
\label{eq:width}
\Gamma_{\oms}=\left[6.2\pm2.1\;(\mathrm{stat.})\pm2.0\;(\mathrm{sys.})\right]\;\mathrm{Me}\kern-.2em\mathrm{V}\!/\!c^{2}.
\end{equation}

\begin{figure*}
\includegraphics[width=19pc]{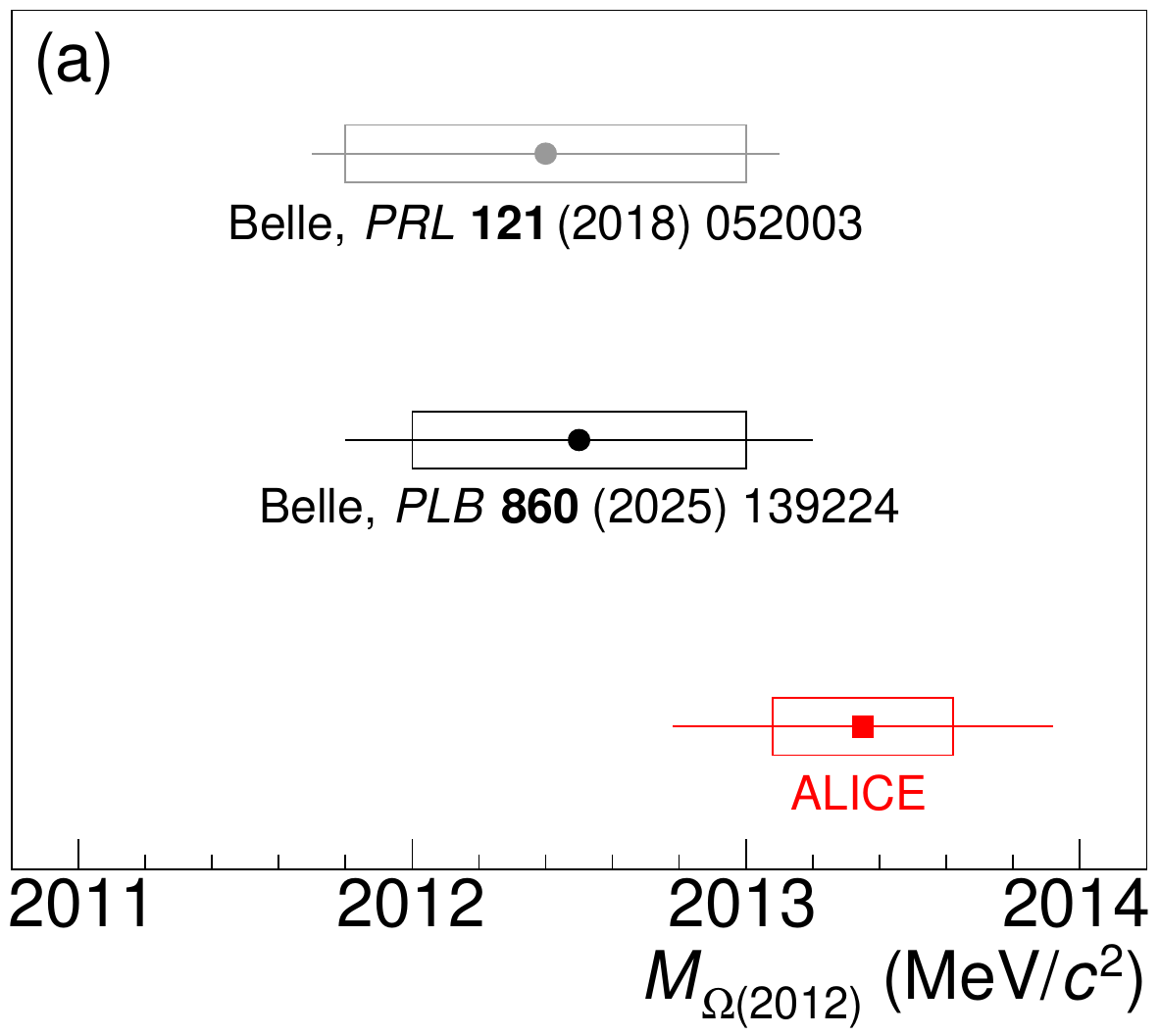}
\includegraphics[width=19pc]{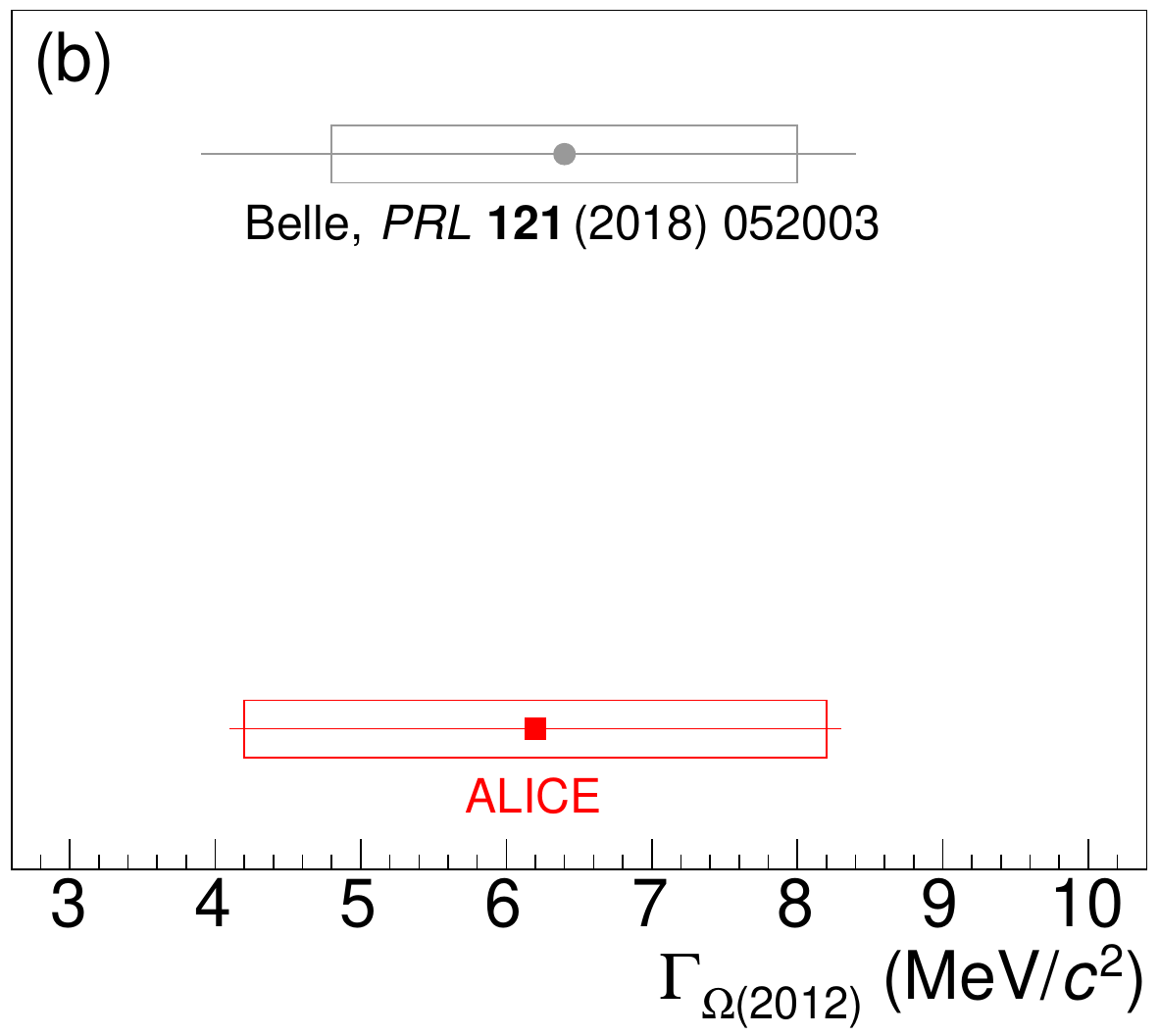}
\caption{\label{fig:comparison}(Color online) Comparison of mass (a) and width (b) measurements of the \oms. The plotted results are from Belle~\cite{Belle_Omega2012_2018,Belle_Omega2012_2025} and ALICE (this work). Statistical uncertainties are shown as bars, and systematic uncertainties are shown as boxes.}
\end{figure*}

Figure~\ref{fig:comparison} shows comparisons of these values to the measurements by Belle~\cite{Belle_Omega2012_2018,Belle_Omega2012_2025}. The ALICE mass value has smaller uncertainties than the Belle measurements and is consistent with both of them. The ALICE measurement of the \oms width is consistent with the Belle measurement and the two have similar uncertainties.

The significance of the observed \oms signal for $\ptt>2.5$~\gvc is 15$\sigma$. This is evaluated as $S/\sqrt{S+B}$, where $S$ is the number of signal counts measured in the range $2.00<\mxk<2.03$~\gvcc and $S+B$ is the combined number of signal and background counts in the same range (without subtraction of the mixed-event background). The first observation of the \oms, reported by the Belle Collaboration~\cite{Belle_Omega2012_2018}, had a significance of $7.2\sigma$.

\subsection{Yields}

\begin{figure*}
\centering
\includegraphics[width=25pc]{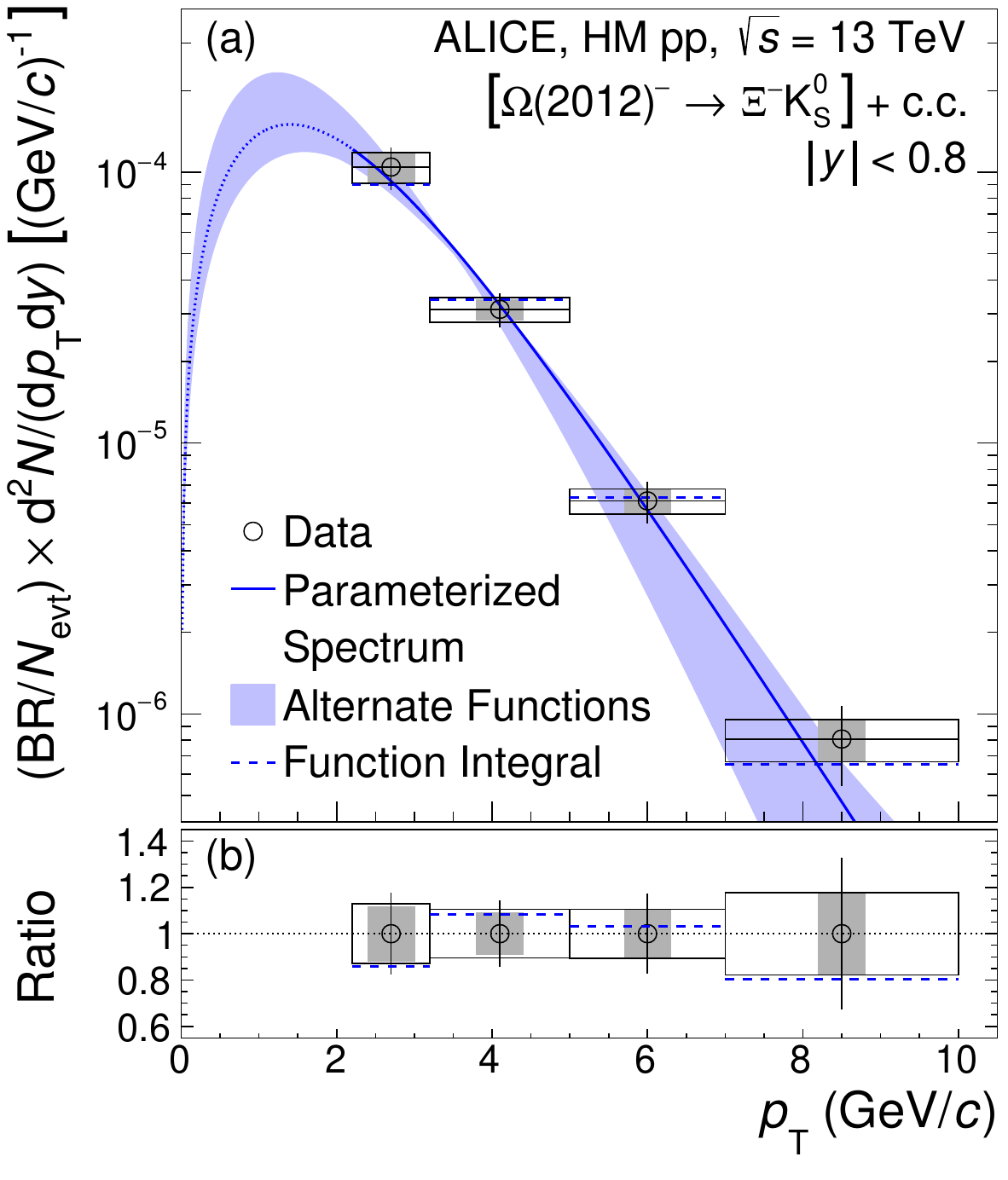}
\caption{\label{fig:spectrum}(Color online) (a) \ptt spectrum of the \oms in HM pp collisions at \rs (not corrected for the unmeasured branching ratio BR for the studied decay channel). Vertical bars represent statistical uncertainties, empty boxes represent the total systematic uncertainties, and shaded boxes represent the portion of the systematic uncertainties that is uncorrelated between \ptt intervals (only slightly smaller than the total systematic uncertainties). The data points are plotted at the center of each \ptt interval and the horizontal bars and boxes span the entire width of each \ptt interval. The curve is the parameterized spectrum for the \oms baryon. The shaded band surrounding the curve indicates the region spanned by the envelope of alternate functions used to describe the spectrum, which affect the extrapolation of the yield to low \ptt. The horizontal dashed lines labeled ``Function Integral'' show the integral of the parameterized spectrum over each of the four measured \ptt intervals. (b) The horizontal dashed lines show the ratio of the parameterized yield to the measured \oms yield in each of the four \ptt intervals. The bars and boxes indicate the fractional uncertainties (statistical, total systematic, and \ptt-uncorrelated systematic) of the measured data.}
\end{figure*}

The \oms transverse-momentum spectrum, not corrected for the unmeasured $\oms\rightarrow\Xi\kz$ branching ratio, is shown in Fig.~\ref{fig:spectrum}. Small additional systematic uncertainties arise from imperfect knowledge of the ALICE material budget and the hadronic interaction cross sections for the decay products; these uncertainties were evaluated as in ref.~\cite{ALICE_strange_pp13} and are listed in Tab.~\ref{table:sys}. The integrated \oms yield for $2.2<\ptt<10$~\gvc (not corrected for the unmeasured branching ratio) is found to be

\begin{equation}
\label{eq:yield_measured}
\mathscr{B}\left[\omsm\rightarrow\Xi^{-}\kz\right]\times\dnodym=\left[1.75 \pm 0.20 \;{\rm (stat.)} \pm 0.15 \;{\rm (sys.)}\right]\!\times\!10^{-4}.
\end{equation}

For the \ptt spectrum, the ``material budget,'' ``hadronic interaction cross section,'' and ``acceptance $\times$ efficiency,'' systematic uncertainties are assumed to be fully correlated between \ptt intervals. The ``decay product selection,'' ``combinatorial background,'' and ``signal extraction'' uncertainties may have both \ptt-correlated and \ptt-uncorrelated components. To determine the \ptt-correlated component, the parameters of the analysis (decay product selections, combinatorial background, fit range, etc.) are varied and the resulting variations in \dnodym are compared to the variations of the yields in the individual \ptt intervals. From this comparison, it is found that the total \ptt-correlated component of these uncertainties is 2.3\% of the yield; the combination of this contribution and the three fully \ptt-correlated uncertainties is 5.1\% on average.

The spectrum is fitted with the aforementioned parameterized \ptt spectrum function, with only the overall scale factor allowed to vary.
The full \ptt-integrated \oms yield is the sum of the yields measured in each of the four \ptt bins, added to the extrapolated yields (derived from the fit function) for the unmeasured low- and high-\ptt regions. The mean transverse momentum \mpt is calculated as follows:

\begin{equation}
\mpt=\frac{\sum\limits_{j=0}^{5}\Big(\mpt_{j}\times Y_{j}\Big)}{\sum\limits_{j=0}^{5}Y_{j}}.
\end{equation}

In the sums, bins 1--4 are the measured \ptt bins, while bins 0 and 5 are the unmeasured low- and high-\ptt regions, respectively. The $Y_{j}$ values are the \oms yields $(\dnody)$ in each bin. The yields $Y_{0}$ and $Y_{5}$ are calculated by integrating the fit function, while the other $Y_{j}$ values are the measured values shown in Fig.~\ref{fig:spectrum} (scaled by the \ptt bin width). The mean transverse momentum in each bin is denoted $\mpt_{j}$; all six values are extracted from the fit function.

Additional systematic uncertainties on these quantities are estimated by using alternate fits to extrapolate the yield; this uncertainty is listed as ``spectrum fit variations'' in Tab.~\ref{table:sys}. The considered variations include the use of different fitting ranges, the alternate parameterized \ptt spectra (obtained from measurements of other baryons as described above), \mT-exponential, Boltzmann, and Fermi-Dirac distributions, and an unconstrained L\'{e}vy-Tsallis function. The full \ptt-integrated \oms yield (not corrected for the unmeasured branching ratio) and the mean transverse momentum are observed to be

\begin{equation}
\label{eq:yield}
\mathscr{B}\left[\omsm\rightarrow\Xi^{-}\kz\right]\times\dnody=\left[4.2 \pm 0.3 \;{\rm (stat.)} ^{+1.5}_{-0.6} \;{\rm (sys.)}\right]\!\times\!10^{-4},
\end{equation}

\begin{equation}
\label{eq:mpt}
\mpt=\left[2.15 \pm 0.08 \;{\rm (stat.)} ^{+0.14}_{-0.26} \;{\rm (sys.)}\right]\;\mathrm{Ge}\kern-.2em\mathrm{V}\!/\!c.
\end{equation}

\subsection{Branching Ratios}

Experimental \oms and $\Omega$ yields can be used together with a statistical model calculation of the $\oms/\Omega$ yield ratio to obtain an estimate of the $\omsm\rightarrow\Xi\overline{\rm K}$ branching ratios. This measurement was obtained for HM pp collisions, but a measurement of the corresponding ground-state $\Omega$ yield has not yet been published. Therefore, the $\Omega$ yield is estimated by interpolation of previous ALICE measurements in pp collisions at \rs at lower multiplicities~\cite{ALICE_strange_pp13} and in \ppb collisions at \rsnn[5.02~\TeV]~\cite{ALICE_Xi_Omega_pPb5}. The $\Omega$ yields appear to follow the same trend as a function of \dnc, independent of collision system or energy. The interpolation is performed using a function of the form

\begin{equation}
\dd N_{\Omega}/\dd y=A\dnc^{n},
\label{eq:extrapolate}
\end{equation}

where $A$ and $n$ are free parameters. The interpolated $\Omega$ yield at $\dnc=31.53$ is estimated to be $0.0179\pm0.0023$. The uncertainty of the interpolated \omx yield originates from the uncertainties of the measured yields, including statistical, multiplicity-uncorrelated systematic, and multiplicity-correlated systematic contributions.

The observed \dnc dependence of the \omx yields in pp and \ppb collisions is well described by statistical thermal models formulated in the canonical ensemble with charge conservation~\cite{Chatterjee_thermal_2017,Sharma_thermal_2019,Vovchenko_2019,Flor_thermal_2022,Cleymans_canonical}. In ref.~\cite{Vovchenko_2019}, analytic parameterizations were obtained to find statistical thermal model parameters as functions of \dnc. For $\dnc=31.53$, these parameterizations give a chemical freeze-out temperature of $T=167$~\MeV, a system volume per unit rapidity of $\dd V/\dd y=75.7\;\mathrm{fm}^{3}$, and a strangeness suppression factor of $\gamma_{\rm S}=0.853$. Since the \omx and \oms have the same minimal quark content, their yields should evolve similarly as a function of \dnc, and any dependence on $\dd V/\dd y$ or $\gamma_{\rm S}$ will cancel. The ratio of the \omx and \oms yields should depend only on the chemical freeze-out temperature of the system, and the masses and spins of the \omx and \oms. The \oms/\omx yield ratio reduces to the thermal density ratio in the grand canonical ensemble~\cite{Hamieh_strangeness_2000}:

\begin{equation}
\frac{\oms_{\rm SM}}{\Omega_{\rm SM}}=\frac{2\mathrm{J}_{\oms}+1}{2\mathrm{J}_{\Omega}+1}\times\frac{M_{\oms}^{2}}{M_{\Omega}^{2}}\times\frac{K_{2}\left(M_{\oms}/T\right)}{K_{2}\left(M_{\Omega}/T\right)}.
\label{eq:ratio_SM}
\end{equation}

Here, $M$ and J denote the masses and spins of the particles, $T$ denotes the chemical freeze-out temperature, and $K_{2}$ is the modified Bessel function of the second type with order 2. For the sake of brevity, here and below, the particle symbols $\Omega$ and \oms are used to represent the particle yields $\dd N_{\Omega}/\dd y$ and $\dd N_{\oms}/\dd y$. Using $T=167$~\MeV, the value for $M_{\oms}$ reported in this paper, and J$_{\oms}=\frac{3}{2}$, it is found that $\oms_{\rm SM}/\Omega_{\rm SM}=0.1666$. Assuming J$_{\oms}=\frac{1}{2}$ scales the yield ratio by a factor of 0.5, while assuming J$_{\oms}=\frac{5}{2}$ scales the yield ratio by a factor of 1.5. The following calculations assume J$_{\oms}=\frac{3}{2}$ for the reasons discussed in the introduction. It has been verified that other thermal-model implementations reproduce the analytic result described above. Using the $T$, $\dd V/\dd y$, and $\gamma_{\rm S}$ values extracted from ref.~\cite{Vovchenko_2019}, the Thermal-FIST framework~\cite{FIST_2019} reproduces the $\oms/\Omega$ yield ratio found using Eq.~\ref{eq:ratio_SM}. A calculation following ref.~\cite{Cleymans_canonical} with $T=156.5$~\MeV also reproduces the analytic result. A $\pm11$~\MeV uncertainty is assigned to the chemical freeze-out temperature. This covers the range of variation in the extracted temperature from recent thermal-model fits of light-flavor hadron yields for collision systems with $\dnc\approx30$~\cite{Chatterjee_thermal_2017,Sharma_thermal_2019,Vovchenko_2019,Flor_thermal_2022,Cleymans_canonical}. This translates into a $\pm13\%$ uncertainty in the $\oms/\Omega$ yield ratio. Using the mass values reported by Belle ~\cite{Belle_Omega2012_2018,Belle_Omega2012_2025} results in a negligible change in this yield ratio.

Combining these results, an estimate of the $\omsm\rightarrow\Xi^{-}\overline{\mathrm{K}}^{0}$ branching ratio is obtained:

\begin{multline}
\mathscr{B}\left[\omsm\rightarrow\Xi^{-}\overline{\mathrm{K}}^{0}\right]= \\ \frac{2\times\mathscr{B}\left[\omsm\rightarrow\Xi^{-}\kz\right]\times\oms_{\rm ALICE}}{\Omega_{\rm ALICE}}\times\frac{\Omega_{\rm SM}}{\oms_{\rm SM}}=0.28^{+0.12}_{-0.07}.
\label{eq:br1}
\end{multline}

The factor of 2 accounts for the fact that only half of the produced $\mathrm{K}^{0}$ and $\overline{\mathrm{K}}^{0}$ mesons decay as \kz. The largest contribution to the uncertainty comes from the ALICE \oms yield, with other independent contributions coming from the uncertainties in the interpolated ALICE $\Omega$ yield and the $\pm10$~\MeV variation in the temperature, which affects the thermal-model calculation of the $\oms/\Omega$ yield ratio.

Belle reported~\cite{Belle_Omega2012_2018} that the ratio of the two-body branching ratios is

\begin{equation}
\frac{\mathscr{B}\left[\omsm\rightarrow\Xi^{0}\mathrm{K}^{-}\right]}{\mathscr{B}\left[\omsm\rightarrow\Xi^{-}\overline{\mathrm{K}}^{0}\right]}=1.2\pm0.3.
\end{equation}

From this, the branching ratio for $\omsm\rightarrow\Xi^{0}\mathrm{K}^{-}$ and the combined two-body branching ratio for $\omsm\rightarrow\Xi\overline{\mathrm{K}}$ are estimated:

\begin{equation}
\mathscr{B}\left[\omsm\rightarrow\Xi^{0}\mathrm{K}^{-}\right]=0.34^{+0.16}_{-0.12},
\label{eq:br2}
\end{equation}

\begin{equation}
\mathscr{B}\left[\omsm\rightarrow\Xi\overline{\mathrm{K}}\right]=0.62^{+0.27}_{-0.17}.
\label{eq:br3}
\end{equation}

The Belle collaboration has measured the three-body $\oms\rightarrow\Xi\pi\overline{\rm K}$ decays~\cite{Belle_Omega2012_2025}. It was observed that these occur predominantly through the resonant channel $\oms\rightarrow\Xi(1530)\overline{\rm K}\rightarrow\Xi\pi\overline{\rm K}$, with no significant non-resonant contribution seen. It was found that the ratio of the three-body branching fraction to the two-body branching fraction is $\mathscr{R}^{\Xi\pi\overline{\rm K}}_{\Xi\overline{\rm K}}=0.99\pm0.26\pm0.06$. Under the assumption that all \oms decays are either $\omsm\rightarrow\Xi\overline{\mathrm{K}}$ or $\omsm\rightarrow\Xi\pi\overline{\rm K}$, the Belle measurement of $\mathscr{R}^{\Xi\pi\overline{\rm K}}_{\Xi\overline{\rm K}}$ would imply a two-body branching ratio $\mathscr{B}\left[\omsm\rightarrow\Xi\overline{\mathrm{K}}\right]=0.50^{+0.08}_{-0.06}$; the ALICE estimate is consistent with this value within uncertainties. Thus, the measured branching ratios reported in Eqs.~\ref{eq:br1}, \ref{eq:br2}, and \ref{eq:br3}, as well as the Belle measurement of $\mathscr{R}^{\Xi\pi\overline{\rm K}}_{\Xi\overline{\rm K}}$, disfavor models of the \oms structure that require large branching ratios (close to unity) for the three-body decays.

The preceding calculations assumed J$_{\oms}=\frac{3}{2}$. If it is assumed J$_{\oms}=\frac{1}{2}$, then the combined two-body branching ratio would have a lower limit of 0.9, which is incompatible with the Belle measurement of $\mathscr{R}^{\Xi\pi\overline{\rm K}}_{\Xi\overline{\rm K}}$. This reinforces the arguments discussed earlier that the \oms is most likely to have a spin of $\frac{3}{2}$.

A cross-check of this branching-ratio study has been performed using measurements and model calculations of the $\Xi(1530)^{0}/\Xi^{-}$ yield ratio (where the particle symbols should be interpreted as also representing the antiparticles). In the ALICE Collaboration's measurement of $\Xi(1530)^{0}$ yields as a function of \dnc in pp collisions at \rs[13]~\TeV, the measured $\Xi(1530)^{0}$ yield was corrected for the $\Xi(1530)^{0}\rightarrow\Xi^{-}\uppi^{+}$ branching ratio~\cite{ALICE_Sigmastar_Xistar_pp13}. That branching ratio was assumed to be exactly $\frac{2}{3}$, based on isospin considerations and value of the total $\Xi(1530)^{0}\rightarrow\Xi\uppi$ branching ratio, which is 100\%~\cite{PDG}. Under this assumption, the $\Xi(1530)^{0}/\Xi^{-}$ yield ratio was reported to be $0.332\pm0.012\;(\mathrm{stat.})\pm0.051$~(sys.) for the highest multiplicity interval reported, with $\dnc=18.67\pm0.20$. Statistical models also assume that the $\Xi(1530)^{0}\rightarrow\Xi^{-}\uppi^{+}$ branching ratio is $\frac{2}{3}$, which is relevant for the calculation of the feed-down of $\Xi(1530)^{0}$ to $\Xi^{-}$. A calculation following ref.~\cite{Cleymans_canonical} gives $\Xi(1530)^{0}/\Xi^{-}=0.36$. If model parameters are extracted based on the parameterizations in ref.~\cite{Vovchenko_2019}, then used with the Thermal-FIST framework~\cite{FIST_2019}, the yield ratio is found to be $\Xi(1530)^{0}/\Xi^{-}=0.38$. The model calculations of the $\Xi(1530)^{0}/\Xi^{-}$ yield ratio, the ALICE measurement of that ratio, and the branching ratio value are all consistent with each other. This indicates that model and experimental yield-ratio values can be employed to estimate unknown branching ratios for other multistrange baryons, as was done above for the \oms.

\section{Conclusion}

In summary, the study reported in this paper corroborates the discovery of the \oms baryon. A signal with a significance of $15\sigma$ has been observed in high-multiplicity-triggered pp collisions at \rs[13~TeV.] The mass and width of the particle are measured to be $\left[2013.35\pm0.57\;(\mathrm{stat.})\pm0.27\;(\mathrm{sys.})\right]$~\mvcc and $[6.2\pm2.1\;(\mathrm{stat.})\pm2.0\;(\mathrm{sys.})]$~\mvcc, respectively; these values are consistent with the previous measurements by Belle~\cite{Belle_Omega2012_2018,Belle_Omega2012_2025}. The first measurement of the \ptt-dependent and \ptt-integrated production yield of the \oms baryon has also been reported, along with a measurement of its mean transverse momentum. Furthermore, based on a comparison to the statistical thermal model expectation, the total branching ratio for the $\omsm\rightarrow\Xi\overline{\mathrm{K}}$ decays is estimated to be $0.62^{+0.27}_{-0.17}$. These results may provide useful information for studies of the structure of the \oms and the chemistry of the matter produced in high-energy hadronic and nuclear collisions. The LHC is in its third running period and the ALICE detector has been upgraded, significantly increasing the number of events it can record. This advancement, combined with implementation of machine-learning techniques to identify \oms candidates, may enable further, more precise measurements of \oms production and properties and may allow for a search for the signatures of chiral symmetry restoration using excited $\Omega$ baryons.


\newenvironment{acknowledgement}{\relax}{\relax}
\begin{acknowledgement}
\section*{Acknowledgements}

The ALICE Collaboration would like to thank all its engineers and technicians for their invaluable contributions to the construction of the experiment and the CERN accelerator teams for the outstanding performance of the LHC complex.
The ALICE Collaboration gratefully acknowledges the resources and support provided by all Grid centres and the Worldwide LHC Computing Grid (WLCG) collaboration.
The ALICE Collaboration acknowledges the following funding agencies for their support in building and running the ALICE detector:
A. I. Alikhanyan National Science Laboratory (Yerevan Physics Institute) Foundation (ANSL), State Committee of Science and World Federation of Scientists (WFS), Armenia;
Austrian Academy of Sciences, Austrian Science Fund (FWF): [M 2467-N36] and Nationalstiftung f\"{u}r Forschung, Technologie und Entwicklung, Austria;
Ministry of Communications and High Technologies, National Nuclear Research Center, Azerbaijan;
Conselho Nacional de Desenvolvimento Cient\'{\i}fico e Tecnol\'{o}gico (CNPq), Financiadora de Estudos e Projetos (Finep), Funda\c{c}\~{a}o de Amparo \`{a} Pesquisa do Estado de S\~{a}o Paulo (FAPESP) and Universidade Federal do Rio Grande do Sul (UFRGS), Brazil;
Bulgarian Ministry of Education and Science, within the National Roadmap for Research Infrastructures 2020-2027 (object CERN), Bulgaria;
Ministry of Education of China (MOEC) , Ministry of Science \& Technology of China (MSTC) and National Natural Science Foundation of China (NSFC), China;
Ministry of Science and Education and Croatian Science Foundation, Croatia;
Centro de Aplicaciones Tecnol\'{o}gicas y Desarrollo Nuclear (CEADEN), Cubaenerg\'{\i}a, Cuba;
Ministry of Education, Youth and Sports of the Czech Republic, Czech Republic;
The Danish Council for Independent Research | Natural Sciences, the VILLUM FONDEN and Danish National Research Foundation (DNRF), Denmark;
Helsinki Institute of Physics (HIP), Finland;
Commissariat \`{a} l'Energie Atomique (CEA) and Institut National de Physique Nucl\'{e}aire et de Physique des Particules (IN2P3) and Centre National de la Recherche Scientifique (CNRS), France;
Bundesministerium f\"{u}r Bildung und Forschung (BMBF) and GSI Helmholtzzentrum f\"{u}r Schwerionenforschung GmbH, Germany;
General Secretariat for Research and Technology, Ministry of Education, Research and Religions, Greece;
National Research, Development and Innovation Office, Hungary;
Department of Atomic Energy Government of India (DAE), Department of Science and Technology, Government of India (DST), University Grants Commission, Government of India (UGC) and Council of Scientific and Industrial Research (CSIR), India;
National Research and Innovation Agency - BRIN, Indonesia;
Istituto Nazionale di Fisica Nucleare (INFN), Italy;
Japanese Ministry of Education, Culture, Sports, Science and Technology (MEXT) and Japan Society for the Promotion of Science (JSPS) KAKENHI, Japan;
Consejo Nacional de Ciencia (CONACYT) y Tecnolog\'{i}a, through Fondo de Cooperaci\'{o}n Internacional en Ciencia y Tecnolog\'{i}a (FONCICYT) and Direcci\'{o}n General de Asuntos del Personal Academico (DGAPA), Mexico;
Nederlandse Organisatie voor Wetenschappelijk Onderzoek (NWO), Netherlands;
The Research Council of Norway, Norway;
Pontificia Universidad Cat\'{o}lica del Per\'{u}, Peru;
Ministry of Science and Higher Education, National Science Centre and WUT ID-UB, Poland;
Korea Institute of Science and Technology Information and National Research Foundation of Korea (NRF), Republic of Korea;
Ministry of Education and Scientific Research, Institute of Atomic Physics, Ministry of Research and Innovation and Institute of Atomic Physics and Universitatea Nationala de Stiinta si Tehnologie Politehnica Bucuresti, Romania;
Ministry of Education, Science, Research and Sport of the Slovak Republic, Slovakia;
National Research Foundation of South Africa, South Africa;
Swedish Research Council (VR) and Knut \& Alice Wallenberg Foundation (KAW), Sweden;
European Organization for Nuclear Research, Switzerland;
Suranaree University of Technology (SUT), National Science and Technology Development Agency (NSTDA) and National Science, Research and Innovation Fund (NSRF via PMU-B B05F650021), Thailand;
Turkish Energy, Nuclear and Mineral Research Agency (TENMAK), Turkey;
National Academy of  Sciences of Ukraine, Ukraine;
Science and Technology Facilities Council (STFC), United Kingdom;
National Science Foundation of the United States of America (NSF) and United States Department of Energy, Office of Nuclear Physics (DOE NP), United States of America.
In addition, individual groups or members have received support from:
Czech Science Foundation (grant no. 23-07499S), Czech Republic;
FORTE project, reg.\ no.\ CZ.02.01.01/00/22\_008/0004632, Czech Republic, co-funded by the European Union, Czech Republic;
European Research Council (grant no. 950692), European Union;
Academy of Finland (Center of Excellence in Quark Matter) (grant nos. 346327, 346328), Finland;
Deutsche Forschungs Gemeinschaft (DFG, German Research Foundation) ``Neutrinos and Dark Matter in Astro- and Particle Physics'' (grant no. SFB 1258), Germany;
ICSC - National Research Center for High Performance Computing, Big Data and Quantum Computing and FAIR - Future Artificial Intelligence Research, funded by the NextGenerationEU program (Italy).
\end{acknowledgement}

\bibliographystyle{utphys}   
\bibliography{refs}

\newpage
\appendix

%
%

\section{The ALICE Collaboration}
\label{app:collab}
\begin{flushleft} 
\small

S.~Acharya\,\orcidlink{0000-0002-9213-5329}\,$^{\rm 50}$, 
A.~Agarwal$^{\rm 133}$, 
G.~Aglieri Rinella\,\orcidlink{0000-0002-9611-3696}\,$^{\rm 32}$, 
L.~Aglietta\,\orcidlink{0009-0003-0763-6802}\,$^{\rm 24}$, 
M.~Agnello\,\orcidlink{0000-0002-0760-5075}\,$^{\rm 29}$, 
N.~Agrawal\,\orcidlink{0000-0003-0348-9836}\,$^{\rm 25}$, 
Z.~Ahammed\,\orcidlink{0000-0001-5241-7412}\,$^{\rm 133}$, 
S.~Ahmad\,\orcidlink{0000-0003-0497-5705}\,$^{\rm 15}$, 
S.U.~Ahn\,\orcidlink{0000-0001-8847-489X}\,$^{\rm 71}$, 
I.~Ahuja\,\orcidlink{0000-0002-4417-1392}\,$^{\rm 36}$
% ...

\section*{Affiliation Notes}

$^{\rm I}$ Deceased\\
$^{\rm II}$ Also at: Max-Planck-Institut fur Physik, Munich, Germany\\
$^{\rm III}$ Also at: Italian National Agency for New Technologies, Energy and Sustainable Economic Development (ENEA), Bologna, Italy\\
$^{\rm IV}$ Also at: Dipartimento DET del Politecnico di Torino, Turin, Italy\\
$^{\rm V}$ Also at: Lehigh University, United States\\
$^{\rm VI}$ Also at: Department of Applied Physics, Aligarh Muslim University, Aligarh, India\\
$^{\rm VII}$ Also at: Institute of Theoretical Physics, University of Wroclaw, Poland\\
$^{\rm VIII}$ Also at: Facultad de Ciencias, Universidad Nacional Aut\'{o}noma de M\'{e}xico, Mexico City, Mexico\\

\section*{Collaboration Institutes}

$^{1}$ A.I. Alikhanyan National Science Laboratory (Yerevan Physics Institute) Foundation, Yerevan, Armenia\\
$^{2}$ AGH University of Krakow, Cracow, Poland\\
$^{3}$ Bogolyubov Institute for Theoretical Physics, National Academy of Sciences of Ukraine, Kiev, Ukraine\\
$^{4}$ Bose Institute, Department of Physics  and Centre for Astroparticle Physics and Space Science (CAPSS), Kolkata, India\\
$^{5}$ California Polytechnic State University, San Luis Obispo, California, United States\\
$^{6}$ Central China Normal University, Wuhan, China\\
$^{7}$ Centro de Aplicaciones Tecnol\'{o}gicas y Desarrollo Nuclear (CEADEN), Havana, Cuba\\
$^{8}$ Centro de Investigaci\'{o}n y de Estudios Avanzados (CINVESTAV), Mexico City and M\'{e}rida, Mexico\\
$^{9}$ Chicago State University, Chicago, Illinois, United States\\
$^{10}$ China Nuclear Data Center, China Institute of Atomic Energy, Beijing, China\\
$^{11}$ China University of Geosciences, Wuhan, China\\
$^{12}$ Chungbuk National University, Cheongju, Republic of Korea\\
$^{13}$ Comenius University Bratislava, Faculty of Mathematics, Physics and Informatics, Bratislava, Slovak Republic\\
$^{14}$ Creighton University, Omaha, Nebraska, United States\\
$^{15}$ Department of Physics, Aligarh Muslim University, Aligarh, India\\
$^{16}$ Department of Physics, Pusan National University, Pusan, Republic of Korea\\
$^{17}$ Department of Physics, Sejong University, Seoul, Republic of Korea\\
$^{18}$ Department of Physics, University of California, Berkeley, California, United States\\
$^{19}$ Department of Physics, University of Oslo, Oslo, Norway\\
$^{20}$ Department of Physics and Technology, University of Bergen, Bergen, Norway\\
$^{21}$ Dipartimento di Fisica, Universit\`{a} di Pavia, Pavia, Italy\\
$^{22}$ Dipartimento di Fisica dell'Universit\`{a} and Sezione INFN, Cagliari, Italy\\
$^{23}$ Dipartimento di Fisica dell'Universit\`{a} and Sezione INFN, Trieste, Italy\\
$^{24}$ Dipartimento di Fisica dell'Universit\`{a} and Sezione INFN, Turin, Italy\\
$^{25}$ Dipartimento di Fisica e Astronomia dell'Universit\`{a} and Sezione INFN, Bologna, Italy\\
$^{26}$ Dipartimento di Fisica e Astronomia dell'Universit\`{a} and Sezione INFN, Catania, Italy\\
$^{27}$ Dipartimento di Fisica e Astronomia dell'Universit\`{a} and Sezione INFN, Padova, Italy\\
$^{28}$ Dipartimento di Fisica `E.R.~Caianiello' dell'Universit\`{a} and Gruppo Collegato INFN, Salerno, Italy\\
$^{29}$ Dipartimento DISAT del Politecnico and Sezione INFN, Turin, Italy\\
$^{30}$ Dipartimento di Scienze MIFT, Universit\`{a} di Messina, Messina, Italy\\
$^{31}$ Dipartimento Interateneo di Fisica `M.~Merlin' and Sezione INFN, Bari, Italy\\
$^{32}$ European Organization for Nuclear Research (CERN), Geneva, Switzerland\\
$^{33}$ Faculty of Electrical Engineering, Mechanical Engineering and Naval Architecture, University of Split, Split, Croatia\\
$^{34}$ Faculty of Nuclear Sciences and Physical Engineering, Czech Technical University in Prague, Prague, Czech Republic\\
$^{35}$ Faculty of Physics, Sofia University, Sofia, Bulgaria\\
$^{36}$ Faculty of Science, P.J.~\v{S}af\'{a}rik University, Ko\v{s}ice, Slovak Republic\\
$^{37}$ Faculty of Technology, Environmental and Social Sciences, Bergen, Norway\\
$^{38}$ Frankfurt Institute for Advanced Studies, Johann Wolfgang Goethe-Universit\"{a}t Frankfurt, Frankfurt, Germany\\
$^{39}$ Fudan University, Shanghai, China\\
$^{40}$ Gangneung-Wonju National University, Gangneung, Republic of Korea\\
$^{41}$ Gauhati University, Department of Physics, Guwahati, India\\
$^{42}$ Helmholtz-Institut f\"{u}r Strahlen- und Kernphysik, Rheinische Friedrich-Wilhelms-Universit\"{a}t Bonn, Bonn, Germany\\
$^{43}$ Helsinki Institute of Physics (HIP), Helsinki, Finland\\
$^{44}$ High Energy Physics Group,  Universidad Aut\'{o}noma de Puebla, Puebla, Mexico\\
$^{45}$ Horia Hulubei National Institute of Physics and Nuclear Engineering, Bucharest, Romania\\
$^{46}$ HUN-REN Wigner Research Centre for Physics, Budapest, Hungary\\
$^{47}$ Indian Institute of Technology Bombay (IIT), Mumbai, India\\
$^{48}$ Indian Institute of Technology Indore, Indore, India\\
$^{49}$ INFN, Laboratori Nazionali di Frascati, Frascati, Italy\\
$^{50}$ INFN, Sezione di Bari, Bari, Italy\\
$^{51}$ INFN, Sezione di Bologna, Bologna, Italy\\
$^{52}$ INFN, Sezione di Cagliari, Cagliari, Italy\\
$^{53}$ INFN, Sezione di Catania, Catania, Italy\\
$^{54}$ INFN, Sezione di Padova, Padova, Italy\\
$^{55}$ INFN, Sezione di Pavia, Pavia, Italy\\
$^{56}$ INFN, Sezione di Torino, Turin, Italy\\
$^{57}$ INFN, Sezione di Trieste, Trieste, Italy\\
$^{58}$ Inha University, Incheon, Republic of Korea\\
$^{59}$ Institute for Gravitational and Subatomic Physics (GRASP), Utrecht University/Nikhef, Utrecht, Netherlands\\
$^{60}$ Institute of Experimental Physics, Slovak Academy of Sciences, Ko\v{s}ice, Slovak Republic\\
$^{61}$ Institute of Physics, Homi Bhabha National Institute, Bhubaneswar, India\\
$^{62}$ Institute of Physics of the Czech Academy of Sciences, Prague, Czech Republic\\
$^{63}$ Institute of Space Science (ISS), Bucharest, Romania\\
$^{64}$ Institut f\"{u}r Kernphysik, Johann Wolfgang Goethe-Universit\"{a}t Frankfurt, Frankfurt, Germany\\
$^{65}$ Instituto de Ciencias Nucleares, Universidad Nacional Aut\'{o}noma de M\'{e}xico, Mexico City, Mexico\\
$^{66}$ Instituto de F\'{i}sica, Universidade Federal do Rio Grande do Sul (UFRGS), Porto Alegre, Brazil\\
$^{67}$ Instituto de F\'{\i}sica, Universidad Nacional Aut\'{o}noma de M\'{e}xico, Mexico City, Mexico\\
$^{68}$ iThemba LABS, National Research Foundation, Somerset West, South Africa\\
$^{69}$ Jeonbuk National University, Jeonju, Republic of Korea\\
$^{70}$ Johann-Wolfgang-Goethe Universit\"{a}t Frankfurt Institut f\"{u}r Informatik, Fachbereich Informatik und Mathematik, Frankfurt, Germany\\
$^{71}$ Korea Institute of Science and Technology Information, Daejeon, Republic of Korea\\
$^{72}$ Laboratoire de Physique Subatomique et de Cosmologie, Universit\'{e} Grenoble-Alpes, CNRS-IN2P3, Grenoble, France\\
$^{73}$ Lawrence Berkeley National Laboratory, Berkeley, California, United States\\
$^{74}$ Lund University Department of Physics, Division of Particle Physics, Lund, Sweden\\
$^{75}$ Nagasaki Institute of Applied Science, Nagasaki, Japan\\
$^{76}$ Nara Women{'}s University (NWU), Nara, Japan\\
$^{77}$ National and Kapodistrian University of Athens, School of Science, Department of Physics , Athens, Greece\\
$^{78}$ National Centre for Nuclear Research, Warsaw, Poland\\
$^{79}$ National Institute of Science Education and Research, Homi Bhabha National Institute, Jatni, India\\
$^{80}$ National Nuclear Research Center, Baku, Azerbaijan\\
$^{81}$ National Research and Innovation Agency - BRIN, Jakarta, Indonesia\\
$^{82}$ Niels Bohr Institute, University of Copenhagen, Copenhagen, Denmark\\
$^{83}$ Nikhef, National institute for subatomic physics, Amsterdam, Netherlands\\
$^{84}$ Nuclear Physics Group, STFC Daresbury Laboratory, Daresbury, United Kingdom\\
$^{85}$ Nuclear Physics Institute of the Czech Academy of Sciences, Husinec-\v{R}e\v{z}, Czech Republic\\
$^{86}$ Oak Ridge National Laboratory, Oak Ridge, Tennessee, United States\\
$^{87}$ Ohio State University, Columbus, Ohio, United States\\
$^{88}$ Physics department, Faculty of science, University of Zagreb, Zagreb, Croatia\\
$^{89}$ Physics Department, Panjab University, Chandigarh, India\\
$^{90}$ Physics Department, University of Jammu, Jammu, India\\
$^{91}$ Physics Program and International Institute for Sustainability with Knotted Chiral Meta Matter (WPI-SKCM$^{2}$), Hiroshima University, Hiroshima, Japan\\
$^{92}$ Physikalisches Institut, Eberhard-Karls-Universit\"{a}t T\"{u}bingen, T\"{u}bingen, Germany\\
$^{93}$ Physikalisches Institut, Ruprecht-Karls-Universit\"{a}t Heidelberg, Heidelberg, Germany\\
$^{94}$ Physik Department, Technische Universit\"{a}t M\"{u}nchen, Munich, Germany\\
$^{95}$ Politecnico di Bari and Sezione INFN, Bari, Italy\\
$^{96}$ Research Division and ExtreMe Matter Institute EMMI, GSI Helmholtzzentrum f\"ur Schwerionenforschung GmbH, Darmstadt, Germany\\
$^{97}$ Saga University, Saga, Japan\\
$^{98}$ Saha Institute of Nuclear Physics, Homi Bhabha National Institute, Kolkata, India\\
$^{99}$ School of Physics and Astronomy, University of Birmingham, Birmingham, United Kingdom\\
$^{100}$ Secci\'{o}n F\'{\i}sica, Departamento de Ciencias, Pontificia Universidad Cat\'{o}lica del Per\'{u}, Lima, Peru\\
$^{101}$ Stefan Meyer Institut f\"{u}r Subatomare Physik (SMI), Vienna, Austria\\
$^{102}$ SUBATECH, IMT Atlantique, Nantes Universit\'{e}, CNRS-IN2P3, Nantes, France\\
$^{103}$ Sungkyunkwan University, Suwon City, Republic of Korea\\
$^{104}$ Suranaree University of Technology, Nakhon Ratchasima, Thailand\\
$^{105}$ Technical University of Ko\v{s}ice, Ko\v{s}ice, Slovak Republic\\
$^{106}$ The Henryk Niewodniczanski Institute of Nuclear Physics, Polish Academy of Sciences, Cracow, Poland\\
$^{107}$ The University of Texas at Austin, Austin, Texas, United States\\
$^{108}$ Universidad Aut\'{o}noma de Sinaloa, Culiac\'{a}n, Mexico\\
$^{109}$ Universidade de S\~{a}o Paulo (USP), S\~{a}o Paulo, Brazil\\
$^{110}$ Universidade Estadual de Campinas (UNICAMP), Campinas, Brazil\\
$^{111}$ Universidade Federal do ABC, Santo Andre, Brazil\\
$^{112}$ Universitatea Nationala de Stiinta si Tehnologie Politehnica Bucuresti, Bucharest, Romania\\
$^{113}$ University of Derby, Derby, United Kingdom\\
$^{114}$ University of Houston, Houston, Texas, United States\\
$^{115}$ University of Jyv\"{a}skyl\"{a}, Jyv\"{a}skyl\"{a}, Finland\\
$^{116}$ University of Kansas, Lawrence, Kansas, United States\\
$^{117}$ University of Liverpool, Liverpool, United Kingdom\\
$^{118}$ University of Science and Technology of China, Hefei, China\\
$^{119}$ University of South-Eastern Norway, Kongsberg, Norway\\
$^{120}$ University of Tennessee, Knoxville, Tennessee, United States\\
$^{121}$ University of the Witwatersrand, Johannesburg, South Africa\\
$^{122}$ University of Tokyo, Tokyo, Japan\\
$^{123}$ University of Tsukuba, Tsukuba, Japan\\
$^{124}$ Universit\"{a}t M\"{u}nster, Institut f\"{u}r Kernphysik, M\"{u}nster, Germany\\
$^{125}$ Universit\'{e} Clermont Auvergne, CNRS/IN2P3, LPC, Clermont-Ferrand, France\\
$^{126}$ Universit\'{e} de Lyon, CNRS/IN2P3, Institut de Physique des 2 Infinis de Lyon, Lyon, France\\
$^{127}$ Universit\'{e} de Strasbourg, CNRS, IPHC UMR 7178, F-67000 Strasbourg, France, Strasbourg, France\\
$^{128}$ Universit\'{e} Paris-Saclay, Centre d'Etudes de Saclay (CEA), IRFU, D\'{e}partment de Physique Nucl\'{e}aire (DPhN), Saclay, France\\
$^{129}$ Universit\'{e}  Paris-Saclay, CNRS/IN2P3, IJCLab, Orsay, France\\
$^{130}$ Universit\`{a} degli Studi di Foggia, Foggia, Italy\\
$^{131}$ Universit\`{a} del Piemonte Orientale, Vercelli, Italy\\
$^{132}$ Universit\`{a} di Brescia, Brescia, Italy\\
$^{133}$ Variable Energy Cyclotron Centre, Homi Bhabha National Institute, Kolkata, India\\
$^{134}$ Warsaw University of Technology, Warsaw, Poland\\
$^{135}$ Wayne State University, Detroit, Michigan, United States\\
$^{136}$ Yale University, New Haven, Connecticut, United States\\
$^{137}$ Yildiz Technical University, Istanbul, Turkey\\
$^{138}$ Yonsei University, Seoul, Republic of Korea\\
$^{139}$ Affiliated with an institute formerly covered by a cooperation agreement with CERN\\
$^{140}$ Affiliated with an international laboratory covered by a cooperation agreement with CERN.\\

\end{flushleft} 

\end{document}